\RequirePackage[2020-02-02]{latexrelease}
\documentclass[twocolumn, switch]{article} % Method A for two-column formatting

\usepackage{preprint}

\usepackage{amsmath, amsthm, amssymb, amsfonts}

\usepackage{natbib}
\usepackage[utf8]{inputenc}	% allow utf-8 input
\usepackage[T1]{fontenc}	% use 8-bit T1 fonts
\usepackage{xcolor}		% colors for hyperlinks
\usepackage[colorlinks = true,
            linkcolor = purple,
            urlcolor  = blue,
            citecolor = cyan,
            anchorcolor = black]{hyperref}	% Color links to references, figures, etc.
\usepackage{booktabs} 		% professional-quality tables
\usepackage{nicefrac}		% compact symbols for 1/2, etc.
\usepackage{microtype}		% microtypography
\usepackage{lineno}		% Line numbers
\usepackage{float}			% Allows for figures within multicol

\usepackage{lipsum}		%  Filler text

\usepackage{newfloat}
\DeclareFloatingEnvironment[name={Supplementary Figure}]{suppfigure}
\usepackage{sidecap}
\sidecaptionvpos{figure}{c}

\usepackage{titlesec}
\titlespacing\section{0pt}{12pt plus 3pt minus 3pt}{1pt plus 1pt minus 1pt}
\titlespacing\subsection{0pt}{10pt plus 3pt minus 3pt}{1pt plus 1pt minus 1pt}
\titlespacing\subsubsection{0pt}{8pt plus 3pt minus 3pt}{1pt plus 1pt minus 1pt}

\usepackage{tikz,xcolor,hyperref}

\definecolor{lime}{HTML}{A6CE39}
\DeclareRobustCommand{\orcidicon}{
	\begin{tikzpicture}
	\draw[lime, fill=lime] (0,0) 
	circle [radius=0.16] 
	node[white] {{\fontfamily{qag}\selectfont \tiny ID}};
	\draw[white, fill=white] (-0.0625,0.095) 
	circle [radius=0.007];
	\end{tikzpicture}
	\hspace{-2mm}
}
\foreach \x in {A, ..., Z}{\expandafter\xdef\csname orcid\x\endcsname{\noexpand\href{https://orcid.org/\csname orcidauthor\x\endcsname}
			{\noexpand\orcidicon}}
}
\title{On the use of {Artificial Neural Networks} in Topology Optimisation}

\usepackage{xwatermark}
\newwatermark[firstpage,color=gray!90,angle=0,scale=0.28, xpos=0in,ypos=-5in]{*correspondence: \texttt{rvawo@dtu.dk}}

\usepackage{authblk}

\author[1\thanks{\tt{rvawo@mek.dtu.dk}}]{Rebekka V. Woldseth\orcidA{}}
\author[1]{Niels Aage\orcidB{}}
\author[2]{J. Andreas B\ae rentzen\orcidC{}}
\author[1]{Ole Sigmund\orcidD{}}

\affil[1]{Department of Mechanical Engineering, Technical University of Denmark. Koppels All\'{e}, B.404, 2800 Kgs. Lyngby, Denmark.}

\affil[2]{Department of Applied Mathematics and Computer Science, Technical University of Denmark. Richard Petersens Plads, B.321, 2800 Kgs. Lyngby, Denmark.}

\usepackage{graphicx}
\usepackage{array}
\usepackage{tabularx}
\usepackage{makecell}
\usepackage{caption}
\usepackage{subcaption}
\usepackage{cprotect}
\usepackage{listings}
\usepackage{textcomp}
\usepackage[framed,numbered]{matlab-prettifier}
\usepackage{appendix}
\usepackage{xurl}

\newcolumntype{L}{>{\raggedright\arraybackslash}X}
\begin{document}

\twocolumn[ % Method A for two-column formatting
  \begin{@twocolumnfalse} % Method A for two-column formatting
  
\maketitle

\begin{abstract}
The question of how methods from the field of artificial intelligence can help improve the conventional frameworks for topology optimisation has received increasing attention over the last few years. Motivated by the capabilities of neural networks in image analysis, different model-variations aimed at obtaining iteration-free topology optimisation have been proposed with varying success. Other works focused on speed-up through replacing expensive optimisers and state solvers, or reducing the design-space have been attempted, but have not yet received the same attention. The portfolio of articles presenting different applications has as such become extensive, but few real breakthroughs have yet been celebrated. An overall trend in the literature is the strong faith in the “magic” of artificial intelligence and thus misunderstandings about the capabilities of such methods. The aim of this article is therefore to present a critical review of the current state of research in this field. To this end, an overview of the different model-applications is presented, and efforts are made to identify reasons for the overall lack of convincing success. A thorough analysis identifies and differentiates between problematic and promising aspects of existing models. The resulting findings are used to detail recommendations believed to encourage avenues of potential scientific progress for further research within the field.
\end{abstract}
%\keywords{First keyword \and Second keyword \and More} % (optional)
\vspace{0.35cm}

  \end{@twocolumnfalse} % Method A for two-column formatting
] % Method A for two-column formatting

%\begin{multicols}{2} % Method B for two-column formatting (doesn't play well with line numbers), comment out if using method A

%%%%%%%%%%%%%%%  Main text   %%%%%%%%%%%%%%%
% \linenumbers
\section{Introduction}\label{sec:0}
Topology Optimisation (TO) is a mathematical approach to mechanical and multiphysics design aimed at maximising structural performance. Spatial optimisation of the distribution of material within a defined domain subject to sets of physical and geometric constraints, effectively increases the design freedom compared to other design approaches. Since the introduction of the homogenisation approach for topology optimisation (\citealt{BendsoeKikuchi1988}), the field has become an increasingly popular academic field as well as a practical design tool for industry, also fuelled by the developments in Additive Manufacturing (AM) permitting production of more complex features to better exploit the increased design freedom gained from TO. While the homogenisation approach demonstrated the promise of TO, it was considered to consist of complex operations and resulted in indistinct blurry optimised designs. Therefore, the SIMP (Solid Isotropic Material with Penalisation) approach (\citealt{Bendsoe1989,ZhouRozvany1991}) soon became the preferred method. This approach considers the relative material density in each element of the Finite Element (FE) mesh as design variables, allowing for a simpler interpretation and optimised designs with more clearly defined features. {Later, alternative approaches to TO emerged, among others, evolutionary algorithms (\citealt{XieSteven1997}), the level-set method (\citealt{Allaireetal2002, Wangetal2003}), feature-mapping methods (\citealt{NoratoBendsoe2004, Guoetal2018, WeinDunningNorato2020}) and stochastic metaheuristics such as Simulated Annealing (SA) and Genetic Algorithms (GA).} The latter non-gradient based TO algorithms have been proven inefficient and intractable for practical problems (\citealt{SigmundGA2011}).\\

Common for the implementation of the different solution methods is that they use an iterative procedure to create a complex mapping from problem-defining characteristics (i.e. supports, loads and objective function) to an optimised structure. To ensure coherence with laws of physics the structure is governed by a system of partial differential equations. As the considered approach to TO is based on nested analysis and design, this system of equations must be solved for the intermediate solution in each iterate of these procedures. For problems increasing in size and complexity obtaining this solution becomes a highly computationally expensive process posing a challenge in large-scale topology optimisation. As accuracy and obtained detail of solutions are highly dependent upon the element size in FE-analysis, the applicability of topology optimisation for real-life design cases is limited by this computational complexity. Therefore, current developments within the field are strongly motivated by the desire to either limit the number of iterations needed to obtain an optimised structure or the computational cost of completing an iteration.\\

The technological development in high-performance computing has not only provided important support in the progress of topology optimisation, but also in other increasingly popular research fields such as Artificial Intelligence (AI). Especially prominent is the growth within the field of Machine Learning (ML) and its subfield Deep Learning (DL), offering promising capabilities in pattern recognition and approximation of complex relations. Machine learning is roughly a collection of model-frameworks for applied function approximations when explicit descriptions of input-target mappings are difficult or impossible (\citealt{Goodfellowetal2018}). {The field has also evolved towards establishing ML-models for approximating probability distributions rather than predicting specific targets (\citealt{LeeGeMa2017}). Due to such developments within the field}, there has been an emergence of algorithms that are able to solve specific tasks, i.e. object or face recognition, better than humans (\citealt{Goodfellowetal2018}). It is especially advances within image-analysis by {Artificial Neural Networks (ANNs) and deep learning} which have motivated the growing research interest in applying such technologies to increase the efficiency of topology optimisation. Assuming a regular finite element mesh is used to discretise the design-domain, one obtains a direct relation between pixels in an image and the material-density distribution throughout the elements in the mesh. This makes the application of well known image-analysis type models directly applicable to the discretised domain in terms of data-representation. Further, as topology optimisation in itself consists of several approximate mappings achieved by iterative solvers, i.e. for solving the PDEs or optimising the sub-problem in each iteration, the idea of reduced and more direct substitutes for these operations is alluring.\\

In the past few years there has been an increase in publications applying AI-frameworks in an attempt to reduce the computational cost of TO. Many of these proposed frameworks are motivated by the resemblance between an element-based material distribution and an image, and the significant developments within deep learning for pattern recognition in image analysis. Increasing attention has, however, not yet yielded much significant progress. The existing literature demonstrates several dead ends, where non-transparent presentations of results oversell the promise of {model architectures} with unrealistic expectations. Neural network models are in some works treated as magic black-boxes with capabilities exceeding human limits, overlooking well-known limitations within the AI field. The idea of iteration-free TO by use of deep learning is particularly prominent, but also problematic. This review article is a reaction to these apparent misconceptions about the current state of AI, and what these models are capable of. Much like \cite{SigmundGA2011} was a response to then current trends in non-gradient TO, this review seeks to clarify why many existing AI-applications in TO seem unfruitful.\\

It is noted that the field of using neural networks for inverse design generally is expanding rapidly, not only in mechanics but also in a wide range of different fields, and not only in TO but also for many other varieties of design parameterisations. This review will mainly concentrate on TO formulations in {solid} mechanics, but also include some discussions about alternative physics applications. Considering the rapid growth and expansion of this field, there is no guarantee that all relevant works are included in this review, however, it is believed that the selection of papers discussed are representative of the current state of the art. \\

The paper is structured as follows; Section \ref{sec:0.2} gives a brief introduction to machine learning {and neural networks}, Section \ref{sec:1} presents the literature considered in this review and the different applications of {neural networks} presented in these articles, Section \ref{sec:2} addresses how to assess and evaluate {such} solution frameworks in TO, Section \ref{sec:666} discusses the limitations of current AI-technology and how these are reflected in the reviewed literature, Section \ref{sec:3} formulates some recommendations for further research into {AI-aided} TO and Section \ref{sec:4} summarises the important findings and comments on future prospects.

\subsection{Artificial Intelligence and {Neural Networks}}\label{sec:0.2}

\textbf{\emph{Artificial intelligence}} is a branch of the computer science field aiming to simulate intelligent behaviour using computers (\citealt{Tiwarietal2018}). The conceptual idea of AI has been present for decades, but the real acceleration in research-interest has only been apparent over the last few years. The resurrection and increasing popularity is a reaction to technological developments improving computing power and techniques, where especially the introduction of GPUs for more efficient parallel processing has been crucial or the determining factor. Currently, AI is one of the hottest research topics due to the prospects of efficient computer-driven applications and the dream of obtaining AI systems capable of matching or succeeding human capabilities. \emph{General AI} refers to the concept of a machine able to mimic the intelligence of humans and can be applied to serve any relevant function. This type of AI is, however, not yet realised and the feasibility of obtaining such machines is unknown. Research, therefore, mostly concerns itself with the area of \emph{Narrow AI}, which is designed for specific applications.\\
Within Narrow AI especially the sub-field of \emph{machine learning}, and subsequently \emph{deep learning}, {has received} increased attention. These sub-disciplines are focused on exploiting existing data to make algorithms or models capable of solving specific problems or serving particular functions.\\

Machine learning refers to the group of methods engineered to complete specific computational tasks intelligently by learning from existing data. The field distinguishes itself from general computational sciences as it aims to automate the task of analytical model building by using data and experience, relieving the degree of human analysis and hardcoded rules needed. This separation between what is seen as human and artificial intelligence is not consistently agreed upon in the scientific community. Some go as far as deeming anything that is programmable as being AI, which would imply that conventional TO is also AI. The authors of this review paper do, however, support the definition presented by \cite{Copeland2016}, which describes the distinction by

\begin{quote}
    \textit{“So rather than hand-coding software routines with a specific set of instructions to accomplish a particular task, the machine is “trained” using large amounts of data and algorithms that give it the ability to \underline{learn} how to perform the task.”}
\end{quote}

A “traditional” gradient-based algorithm is indeed hand-coded and does not have any such built-in learning aspects. All changes and update rules are pre-programmed. Given deterministic computing conditions and perfect arithmetics, repeated applications will arrive at the exact same final solution, even if the algorithm navigates through a complicated design space. Potential variations in final solutions may be caused by imperfect arithmetics or non-deterministic computing, i.e. due to parallel execution, but these variations are not deliberate actions the algorithm does to improve performance for the next run, and hence nothing is learned. The same can be said about genetic algorithms. Given the same starting conditions and random seed, the algorithm will always converge to the same solution. If later solving a slightly perturbed design problem, there is no mechanism for exploiting knowledge from the previous study to improve the solution obtained for the new problem. With this definition, it is thus not correct to categorise TO in its traditional form as AI.\\

To illustrate what actually does qualify as AI, based on the presented definition, this section aims at introducing the core concepts of machine learning relevant for topology optimisation. The methods applied in the papers reviewed are specialised models based on versions and combinations of those to be presented in this brief theoretical introduction.\\

%Most ML-methods used for topology optimisation in the literature are concerned with the sub-field of \emph{Artificial Neural Networks} (ANNs).
{The majority of ML-methods used for topology optimisation are deep learning frameworks, meaning they are based on the use of \emph{Artificial Neural Networks} (ANNs). Therefore, the following theory will focus on introducing such ANN-based methods.} This family of methods is popular due to the associated design-flexibility resulting in possible modifications for a wide variety of applications (\citealt{Janieschetal2021}). An ANN mimics information processing in biological systems by modelling connected processing units referred to as neurons, where the connections between them represent signal transmissions.\\

\begin{figure}[ht]
    \centering
    \includegraphics[width=\linewidth]{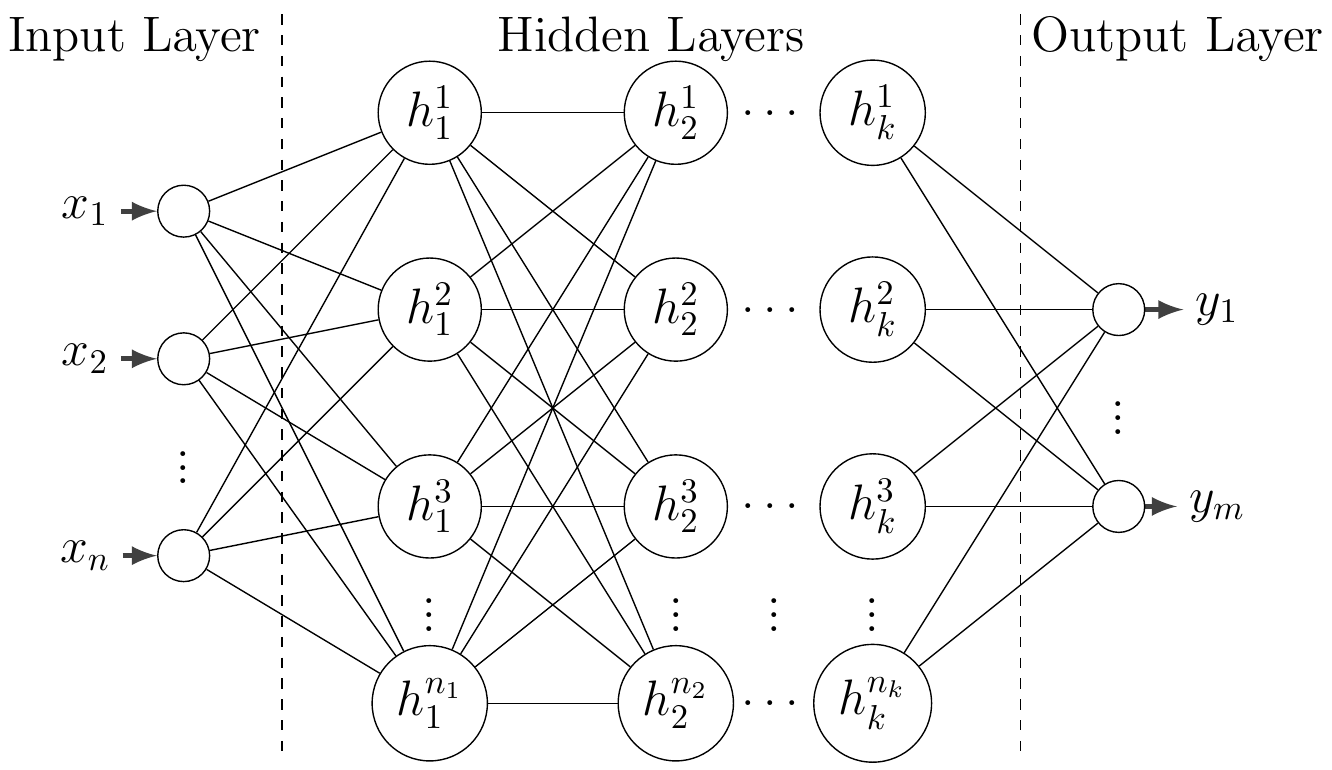}
    \caption{General structure of an ANN illustrating how the input-array $\boldsymbol{x}$ flows through the layers of the network and is translated to some output-array $\boldsymbol{y}$. Different network architectures are achieved by varying the number of hidden layers, the number of nodes within each hidden layer and the connections between the nodes in different hidden layers.}
    \label{fig:ANN_general}
\end{figure}

In simple terms ANNs are used to represent non-linear functions, mapping some $n$-dimensional input to some $m$-dimensional output. Fig. \ref{fig:ANN_general} illustrates the general structure of an ANN where the neurons are represented by nodes and the signal transmissions as edges. The network consists of three elementary types of layers, namely input, hidden and output layers. Data is passed to the network in the input layer, and passed on through the hidden layers using the neural connections, before the mapped result is passed to the output layer. During each connection the signal from the origin node is multiplied by a weight and added to a bias before being passed through an activation function associated with the destination node. Different activation functions may be used for separate parts of the network, and the general definition is that such a function determines how the weighted sum of the neural input is transformed to the appropriate neural output. The typical choices of  such activation functions is what introduces non-linearity in the ANN.\\

The different layers in an ANN as such represent nested function evaluations of the input data to obtain the desired output format. The nature of the overall model is determined by the network architecture in terms of number of hidden layers and number of neurons associated with each of these layers, as well as the weights, biases and activation functions used. The weight and bias parameters of a network are determined through the training process, where the model is fitted to the desired application based on available data. Training an ANN can be seen as a form of complex regression analysis or an optimisation problem, aimed at obtaining {the best cost function for the model based on the desired input-output characteristics. Depending on the specific application this cost function may include simple measures like prediction accuracy or more complex measures such as distributional transport or equilibria to min-max games as discussed further below.} There are several different learning algorithms available for such tasks, these are typically categorised by the characteristics of the desired model and the available data for knowledge extraction.\\
\begin{figure*}[ht]
    \centering
    \includegraphics[width=\linewidth]{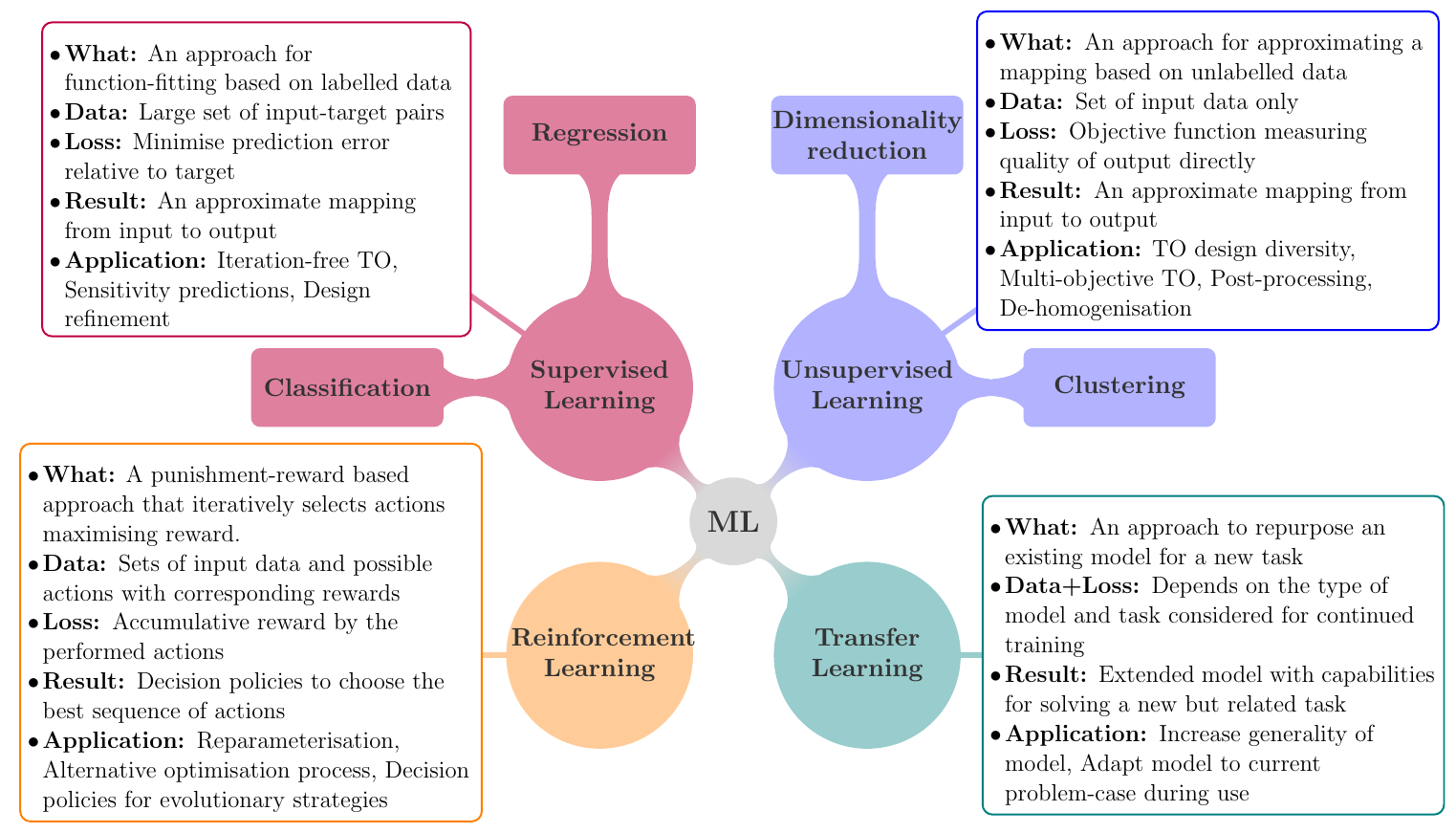}
    \caption{An overview of the common learning strategies {from ML used for ANNs} and their respective areas of application in TO.}
    \label{fig:ML_infographic}
\end{figure*}
{Figure} \ref{fig:ML_infographic} gives a general overview of {learning methods} in terms of training strategy and how they relate to TO applications. Supervised and unsupervised learning constitute the most commonly considered strategies for fixed datasets while reinforcement learning is a different experience-based approach (\citealt{Goodfellow2016}). Transfer learning acts as an extension applicable to any of the other strategies. To further elaborate on the fundamentals of ML it is also relevant to give an introduction to the mathematical models commonly applied within each of these categories.\\

\textbf{\emph{Supervised learning}} is used when the training data consists of inputs with known corresponding output values, i.e. the ${x}$-values and desired $y$-values of Fig. \ref{fig:ANN_general} are known for each data sample. In this case the parameters of the network are updated as to minimise the prediction error, i.e. the \emph{{loss function}}, between the model-obtained ($\boldsymbol{x}$) and target output ($\boldsymbol{y}$) across each of the training samples, such that an approximate map from input to output is achieved. Supervised learning is commonly applied when the aim is to achieve iteration-free TO (\citealt{Abueiddaetal2020, YuHurJungJang2019}) or when an efficient approximation of sensitivities is desired (\citealt{QianYe2020}). In the first case the inputs could be the problem boundary conditions and applied loads while the outputs are {corresponding pre-optimised} structures. In the latter case the inputs also include information about the structural design, i.e. the element density values, and the outputs could for instance be the displacement field or strain energy density for each element in the structure.\\

\textbf{\emph{Unsupervised learning}} is used when the model should detect underlying patterns without any predefined output images. {Instead, the formulation of the loss function alone controls the objective of the learning process. Care is therefore needed to ensure that the loss function measures the performance for the intended task, accounting for all aspects of what defines a desired output. On the other hand, this feature makes unsupervised learning more suited than supervised learning for problems where there are multiple useful outputs for each input. This would, however, require that one is only interested in one of the possible outputs for each input and that there exists an appropriate function for measuring the quality of the output. The unsupervised approach to achieving iteration-free TO could therefore circumvent the generation of pre-optimised structures by including the compliance and volume fraction constraint in the loss function (\citealt{Halleetal2020}). Unsupervised learning can also be used directly as an optimisation process of a reparameterised design representation (\citealt{ChandrasekharSuresh2020, DengTo2021}), or post-processing of an optimised structure by de-homogenisation (\citealt{Elingaard2021}).}\\

{\textbf{\emph{Reinforcement learning}} is a process for discovering policies for how to best choose a sequence of actions to evolve a system from an initial state to reach some predefined goal. The loss function equivalent for this learning procedure is defined by a reward-punishment scheme evaluating the quality of an action. Reinforcement learning differs from unsupervised learning in that the possible system states and actions must be pre-defined. This strategy is useful for conducting optimisation tasks which can be reformulated as a Markov decision process like binary optimisation of trusses by evolutionary strategies (\citealt{HayashiOhsaki2020}). In this case, the system states correspond to a truss structure, formed by a set of members, the actions are the removal of some structural members, and the punishment or reward is measured by whether the chosen action leads to a violation of constraints (i.e. compliance or stress). As such, the goal is to achieve an optimised structure satisfying the constraints by iteratively removing structural members, and the trained model is used to determine what structural members to remove at each iteration. This learning strategy can also be used for exploration of the design space by choosing different parameter settings for topology optimisation (\citealt{SunMa2020,JangYooKang2020})}\\

{\textbf{\emph{Transfer learning}} refers to when a pre-trained model developed to solve a specific task is re-purposed to a second task by using the parameters and biases of the pre-trained network as initial settings in the training process for the new task.} Knowledge gained from training the model to handle the initial task is as such used to limit the effort in terms of data samples and computational time needed to obtain good performance for solving a different but related task. The applicability and viability of transfer learning heavily depends on the generality of the initial task or how closely the different tasks are related. {Transfer learning could be used to improve the performance of a model trained for iteration-free TO on new problems with boundary conditions, length scale or constraints different from those covered in the training cases for the original model.}\\

Several of the solution frameworks contained in this review incorporate what is deemed \emph{active} or \emph{online} learning. This could either indicate that transfer learning is conducted during the optimisation procedure to improve the performance for the specific problem being solved, or that the learning procedure is re-started for each instance, where sequential transfer learning ensures adaption to the current problem. As for transfer learning, both supervised and unsupervised learning can be utilised in this way.\\

Based on some initial guesses for the weights and biases, an optimisation procedure is used to iteratively update the parameters to improve the performance measured by the {loss function}. There are different learning algorithms with different settings available to perform this training process, and an appropriate choice of method and settings is up to the designer. \\

In supervised and unsupervised learning some version of a gradient-based optimisation algorithm like {gradient descent} and Levenberg-Marquardt is commonly employed to determine the best choice for model parameters. The concept of backpropagation is often exploited to compute the desired gradients. \emph{Backpropagation} simply refers to the procedure of computing the gradient of the {loss function} with respect to the model parameters using the chain rule, very similar to TO with multiple filtering operations (\citealt{WangLazSig11}). Reinforcement learning problems are represented as Markov Decision processes and {differ from} the other training categories in that the sequence of actions chosen are dependent. Therefore, the frameworks of the applied training algorithms are inspired by Dynamical Programming or probabilistic methods like Monte Carlo simulation. \\

In addition to the training procedure setup, the activation functions and the network architecture must be determined. The network architecture is defined by the number of hidden layers, the number of nodes in each of these layers and the connectivity between nodes of different layers. These characteristics are commonly referred to as the network hyperparameters and can both be determined manually or by a separate optimisation-like routine (\citealt{Goodfellow2016}). \\

The number of hidden layers {is} used to distinguish between shallow and Deep Neural Networks (DNNs). It is especially within the deep learning segment that technological advances have had an important influence, as an increasing number of hidden layers and thus an increasing model complexity, requires both more robust learning algorithms and more efficient hardware technology. As the number of parameters in a network grows, so does the memory consumption and time needed for training. Therefore it is crucial to exploit the flexibility in the network structure to increase performance for the desired task, while limiting the size of the network.\\

There are some well-established network architecture types that form the foundations for most ANN models. \emph{Feedforward NNetworks} (FNN) or \emph{Multilayer Perceptrons} (MLPs) are acyclic ANNs where information only moves forward from the input layer, through the hidden layers sequentially, towards the output layer (\citealt{Goodfellow2016}). Feedforward neural networks are usually used for supervised learning of data that is neither sequential nor time-dependent. A network is fully connected if each neuron is connected to all neurons in the next layer. Such networks are useful as no special assumptions need to be made about the structure of the input. The drawback is however that this generality may hamper the model performance and require  unnecessarily high computational costs.\\

\textbf{\emph{Convolutional Neural Networks}} (CNNs) are a special case of feedforward networks which are not fully connected, and where weight sharing is used to make the networks translation equivariant. Translation equivariance means that the network has the same output for given features, regardless of where they are located in the input. CNNs are particularly useful for treating regular grids such as 2D or 3D images (\citealt{Janieschetal2021}). A crucial ingredient in a CNN is the use of an ANN as a filter which can be seen as a sequence of discrete convolutions where each is followed by a non-linear mapping. Considering the input image to be the discrete function being convolved, the network weights define the {convolution function} while the activation functions introduce the non-linearity in the network. The concept of weight sharing means that the same filter can be placed in different locations of the input image reusing the same weights to extract the same features. Typically, many layers of such filters are used, paired with pooling, downsampling, or upsampling between layers. The term CNN refers to the entirety of the network constructed by these multiple layers of filters. Note that the weights are not shared between layers but only within each layer.\\
As such, weight sharing allows for the network to be trained to recognise the same objects anywhere in the image, even if the object placement is not varied in the training dataset. Another important benefit of weight sharing is that the size of the network is reduced, in terms of the number of parameters one need to adjust during training. The CNN-architecture therefore allows for using fewer training data samples to create a smaller network with improved performance.\\

The mentioned network architectures are suitable for supervised training and generating \emph{discriminative models}, which are usually used for regression or classification with known output features. Alternatively, there are \emph{generative models} that aim to learn some data distribution through unsupervised learning. One such model is the \emph{Variational Autoencoder} (VAE) which aims at learning how to efficiently represent the data by compressing it to a latent vector, and consequently how to translate from such a latent vector back to the original input format. The corresponding network thus has an encoder-decoder structure similar to CNN, but the purpose is to accomplish a proficient dimensionality reduction of the data. Based on a trained VAE, new data instances can then be generated by sampling in the latent space and subsequently applying the decoding procedure. {This allows for training a VAE to reduce the dimensionality of the design representation such that optimisation can be performed by iteratively updating the latent vector (\citealt{Guoetal2014}).} The drawback of the VAE is that when used to generate new data one can obtain blurry samples as a consequence of the learned average data representation.\\
\emph{Generative Adversarial Networks} (GANs) take a different approach to the generative task by coupling a generator network with a discriminator used to judge the quality of the data samples created by the generator. The training of a GAN constitutes a min-max game between the two networks where the generator aims at improving its ability to create {``}fake'' data samples imitating the available training data, while the discriminator is trained to distinguish whether some input data sample is a generated fake or not. By this procedure the GAN learns how to create new seemingly real data samples. The provided {data samples} could consist of TO optimised structures, where the generator then learns to generate new believable structural layouts and the discriminator learns how to detect inappropriate structures. {In application} the provided data samples could consist of TO optimised structures, where the generator then learns to generate new believable structural layouts and the discriminator learns
how to detect inappropriate structures. {The model can then be used within a framework for diversifying design options for a specific mechanism (\citealt{OhJungKim2019,RawatShen2018})}\\
{A third family of generative models is \emph{Normalising Flows} (NFs) which, in contrast to VAEs and GANs, explicitly learns the probability density function of the input data (\citealt{Kobyzev2021}). These models are constructed by invertible transformations mapping the complex distribution of observed data to a standard Gaussian latent variable. The latent space is in this case of the same dimensionality as the input and does therefore not suffer from the loss of information by averaging as VAEs do, where the latent space commonly serves as a compression of the input data. The invertible nature of the NF models allows for loss-less reconstruction of the input data and generative potential. This generative potential allows for high-dimensional image (\citealt{Kingma2018,Dinh2016}) and point cloud (\citealt{YangHuangHao2019}) generation and could be utilised for TO in a similar manner to VAE and GANs. NFs can further be used to map between image and point-cloud representations (\citealt{Pumarola2020}) which could indicate potential for post-processing of TO optimized structures. To the best of our knowledge, there are no works in the current literature utilising NFs for TO.}\\

As such, an overview of the modelling principles at the core of ML-applications for TO has been presented. How these strategies are combined and exploited to aid in the development of TO solution frameworks is covered in more detail by the following literature review.

\section{Literature Review}\label{sec:1}
Initially the motivation behind combining AI with TO was related to the increasingly successful utilisation of deep-learning models for image analysis and generation. This is reflected within the currently most popular applications of AI in TO, where some NN-architecture is trained in the hopes of generating viable structural images given problem-descriptive inputs. In such approaches, one seeks to develop an AI-methodology replacing the need for conventional iterative optimisation methods. Other applications of AI-methods related to sub-procedures of the optimisation process are however also receiving increasing interest, with the hopes that one can develop models to support or fully replace certain computationally expensive components of the solution procedure.

\subsection{Overview}\label{sec:1.1}
The current literature on AI in TO can be categorised into five main groups. For the purpose of this review, these categories are defined as \emph{Direct design}, \emph{Acceleration}, \emph{Post-processing}, {\emph{Reduction} and \emph{Design diversity}}. This section will give a quick overview over what these categories entail and connect them to the principal AI concepts utilised within each category. \\

\emph{Direct design} refers to the strategy of creating {learning} models to directly predict an optimal structure when given some problem descriptive characteristics, and as such the aim is to achieve optimal structures ``instantly'', in an iteration-free manner.\\

\emph{Acceleration} refers to {learning} models used as supplements to conventional iterative solution methods, with the aim of reducing the computational costs. This is typically achieved through replacing the FE analysis with some approximate model at a subset of the iterations, or by constructing a direct mapping between intermediate structures effectively skipping some subset of iterations.\\

\emph{Post-processing} is defined as the modification of structures obtained through conventional TO or homogenisation usually aimed at ensuring manufacturability by changing the shape, determining microstructure configurations, smoothing of boundaries or as a substitute for de-homogenisation approaches.\\

\emph{Reduction} is performed with the aim of reducing the size of the design space by constructing a model that describes the topology in a more compact way. This reparameterisation then allows for iterative optimisation with fewer design variables, which effectively speeds up the solution procedure. {{Note} that such approaches resemble standard Model Order Reduction methods, but with the distinction that the nature of the AI approaches is different, since these are not explicitly programmed.}\\

\emph{Design-diversity} concerns generating multiple design solutions to the same topology optimisation problem and is somewhat related to finding the Pareto-front in multi-objective optimisation. A set of several candidate structures exhibiting different desired characteristics are generated providing multiple different design options to choose from.\\

For ease of describing trends within the different application areas Table \ref{tab:categories} sorts most of the reviewed articles into appropriate, more specific sub-categories of each of the five main groups. 
\begin{table*}[ht]
\centering%
\caption{An overview of how {82 out of 111} articles in this review distributed amongst the different application-based categories.}
\label{tab:categories}
\begin{tabularx}{\linewidth}{l|LLL}
\toprule
        \textbf{{Direct Design}} &  \multicolumn{3}{>{\hsize=\dimexpr3\hsize+3\tabcolsep+\arrayrulewidth\relax}X}{\cite{Abueiddaetal2020}, \cite{AtesGorguluarslan2021}, \cite{Behzadi_Ilies2021}, \cite{BehzadiIllies2021}, {\cite{Bieleckietal2021}}, \cite{CanYaoRen2019}, \cite{GarreltsHuberetal2021}, \cite{Halleetal2020}, \cite{Harishetal2020}, \cite{Herath2021}, {\cite{Hoangetal2022}}, \cite{Leietal2019}, \cite{LiBetal2019}, \cite{LiKirbyZhe2020}, \cite{LuoZhouetal2021}, \cite{MaZeng2020}, \cite{NakamuraSuzuki2020}, \cite{Nieetal2020}, {\cite{Radeetat2020}}, \cite{UluZhangKara2016}, \cite{WangXiangetal2021}, \cite{YanZhangXuetal2022}, \cite{YuHurJungJang2019}, {\cite{ZhengFanetal2021}}, \cite{ZhengHeLiu2021} }\\\midrule
        \textbf{Acceleration} %& Function evaluations:, 
        &\multicolumn{2}{>{\hsize=\dimexpr2\hsize+2\tabcolsep+\arrayrulewidth\relax}X}{Sensitivity analysis: \cite{AuligOlhofer2013}, \cite{AuligOlhofer2014}, \cite{AuligOlhofer2015}, {\cite{Barmadaetal2021}}, \cite{QiuDuYang2021}, \cite{ChiZhangetal2021}, \cite{KesKirNara2021}, \cite{Lee_etal2020}, \cite{Papadrakakis1998}, \cite{QianYe2020}, \cite{Sasaki2019},\cite{YueYangDuetal2021}, \cite{ZhangChiPaulinoetal2021} } 
        & {Convergence: \cite{Bangaetal2018}, \cite{JooYuJang2021}, \cite{Kallioras2020}, \cite{KalliorasLagaros2020}, \cite{KalliorasNordasLagaros2021}, \cite{KesAliTas2021}, \cite{Linetal2018}, \cite{SosnovikOseledets2017}, \cite{Xueetal2021}, \cite{YeLietal2021}}
        \\\midrule
        \makecell[l]{\textbf{Post-}\\\textbf{processing}} & {Shape optimisation: {\cite{Hertleinetal2021}}, \cite{LinLin2005}, \cite{Vulimirietal2021}, \cite{Yildizetal2003} } %Manufacturability
        &\multicolumn{2}{>{\hsize=\dimexpr2\hsize+2\tabcolsep+\arrayrulewidth\relax}X}{Upscaling: \cite{Elingaard2021}, \cite{LiBetal2019}, \cite{Napieretal2020}, \cite{Wangetal2020}, \cite{Xueetal2021}, \cite{Yooetal2021}}
       % & {Microstructure configuration: \cite{Wangetal2020} }
        \\\midrule %& Microstructure configuration
        \textbf{Reduction} & \multicolumn{3}{>{\hsize=\dimexpr3\hsize+3\tabcolsep+\arrayrulewidth\relax}X}{\cite{ChandrasekharSuresh2020}, {\cite{ChandraSuresh2021}}, \cite{ChandraSuresh2022}, {\cite{ChenShen2021}}, \cite{DenTo2020}, \cite{DengTo2021}, \cite{Guoetal2018}, \cite{Greminger2020}, { \cite{HayashiOhsaki2020},} \cite{Hoyeretal2019}, \cite{Zehnderetal2021}, {\cite{ZhangZhao2021}}, \cite{ZhuGuoetal2021} } 
        \\\midrule
        \makecell[l]{\textbf{Design}\\ \textbf{Diversity}} &  \multicolumn{3}{>{\hsize=\dimexpr3\hsize+3\tabcolsep+\arrayrulewidth\relax}X}{\cite{JangYooKang2020}, \cite{KesBidKel2020}, \cite{OhJungKim2019}, \cite{RawatShen2018}, \cite{RawatShen2019}, {\cite{RawShe2019}}, {\cite{ShenChen2019}}, \cite{SunMa2020}, {\cite{Satoetal2019}}, \cite{Yamasakietal2021}  }
        \\ \bottomrule
\end{tabularx}
\end{table*}

\subsection{Categorisation}\label{sec:1.2}
Within each of the five main categories presented, the research is {built on} similar fundamental ideas and motivations. Further, the resulting model performances exhibit mostly comparable strengths and weaknesses. Therefore, this section will focus on the contents of each category in a collective manner, highlighting works if distinction is deemed necessary.

\subsubsection{Direct design}
The direct design model approach is currently one of the most popular applications of AI in TO, and the aim is to directly achieve an optimised structure for a given problem definition, completely removing the need for expensive iterative procedures. Commonly this is achieved by implementing {neural network architectures} popular in image segmentation, like CNN or GAN. The structural design representation is typically defined by element densities within a regular FE-mesh, similar to the conventional SIMP approach, but some base their structural representation on geometrical features inspired by {Feature Mapping or Moving Morphable Components (MMC) techniques (\citealt{ZhengFanetal2021, Hoangetal2022})}.\\
The considered optimisation problem is usually minimum compliance subject to a volume constraint, but other applications like thermal conduction problems, \cite{LiBetal2019} and \cite{Linetal2018}, are also considered. Model inputs consist of boundary conditions, applied forces and volume fraction, given in spatial representation by a sequence of {input matrices} with dimensions equal to those of the considered FE-mesh. In certain works, additional inputs related to initial stress or strain (\citealt{Nieetal2020,YanZhangXuetal2022}) and displacement fields (\citealt{WangXiangetal2021}) are also included. The trained network is then used to map these inputs to some final structural design, either by regression as continuous {grey scale} element-density values, or by classification as binary black-and-white values indicating whether material is present within an element or not. \cite{GarreltsHuberetal2021} presented a slightly different approach aiming at training a model to also handle rotated pictures taken of hand-sketched boundary conditions as input, and then mapping this image to an optimised Michell structure.\\

%\textcolor{red}{I have changed the order and some of the formulations in the rest of this section, is this better?}\\
Most of the direct design models are trained in a supervised (CNN) or semi-supervised (GAN) manner where a large number of optimised structures are used as training output target samples. This means that at least one complete run of conventional TO must be completed for each problem case considered in the training process. Therefore it is critical to limit the size of the needed training dataset as well as the number of elements in each sample to make the computational time for building the desired model viable. This is however in conflict with obtaining a well-performing model that is able to handle a wide variety of different problems, as neural networks perform better on inputs that are similar to the previously seen training samples. These factors are likely the reason for most direct design models in the literature focusing on problems with fixed or very similar support conditions, only varying the volume fraction and applied loads, where the number of and possible placements of loads typically also come with limitations.\\

Common for most of these network architectures, is that they avoid fully connected layers. Therefore, in theory, they allow for flexibility in terms of the dimensions of and number of elements in the considered FE-mesh. Still, the presented training and test problems are typically restricted to a small fixed mesh where the {input matrices} and the output image for the network are explicitly defined by the dimensions of this fixed mesh, such that this potential adaptability is not exemplified. Further, when the network is only trained for the same regular {mesh dimensions} as a direct mapping from boundary conditions to an optimised structure, it is unclear whether the model is readily translated to problems with different {mesh dimensions} or resolutions, even given the inherent flexibility of the CNN. \cite{ZhengHeLiu2021} made some effort to ensure {mesh flexibility} by designing a network for a larger reference mesh with a mechanism for defining empty elements, such that the {mesh dimensions} could be varied within this reference domain. Their approach implies that the fixed-dimension reference mesh is defined before training, where it must be large enough to encompass the meshes of all problems the network will be used to solve in the future. In this review, it is found that none of the published direct design models explore whether it is possible to fully exploit the generality offered by CNN-types networks, in terms of input image dimensions.\\

If the {network architecture} is designed specifically for certain mesh dimensions, then the network size increases with the number of elements in the mesh, resulting in more parameters to be determined during training and a larger memory consumption for storing the model. In turn, this also increases the cost of obtaining training samples, as finer meshes imply more time needed to optimise a structure using conventional TO. It appears that such perceived {mesh dependency} of the direct design models is a limiting factor for research within this class of approaches. Further, the expense of generating high-resolution training samples and the increased structural complexity associated with higher resolution FE-meshes, are likely determining factors explaining why most of the current literature only considers low-resolution meshes, {typically with fewer than 4,000 elements and at the most 26,000 elements (\citealt{Leietal2019,LiBetal2019,ZhengHeLiu2021}), which is five orders of magnitude below state-of-the-art TO methods using two billion elements (\citealt{Baandrup2020})}.\\

\cite{YuHurJungJang2019} considered 2D coarse grid problem cases with fixed boundary conditions, randomly sampled volume fraction between 0.2 and 0.8, and random single-point directional force application. By repeated sampling and application of open-source topology optimisation code (\citealt{Andreassenetal2011}) 100,000 corresponding optimised structures are generated, where a random subset of 80,000 of these are used for training and validating the network, while the remaining 20,000 are used for testing. The restricted sampling space, the large number of generated structures and the random selection of training and test data means that there is a high likelihood of each test-sample being similar to one of the training cases. Still, the reported results show that the prediction ability of the model is lacking when applied to the test-cases as larger structural disconnections are apparent in the predicted structures. Thus, thousands of expensive datasamples are collected to train a network which fails at solving problems strongly related to those seen by the network during training. \cite{NakamuraSuzuki2020} used the results reported by \cite{YuHurJungJang2019} as a benchmark for their direct design network. By increasing the number of optimised structures used for training and validation of the model to 330,000, within the same sample-space, they reported a greater prediction accuracy in terms of pixel-wise density errors, as expected when allowing for more than three times the number of training instances. However, the worst case solutions still exhibit structural disconnections, implying a large prediction error in terms of the compliance of the design.\\

\begin{figure}[ht]
    \centering
    \includegraphics[width=0.75\linewidth]{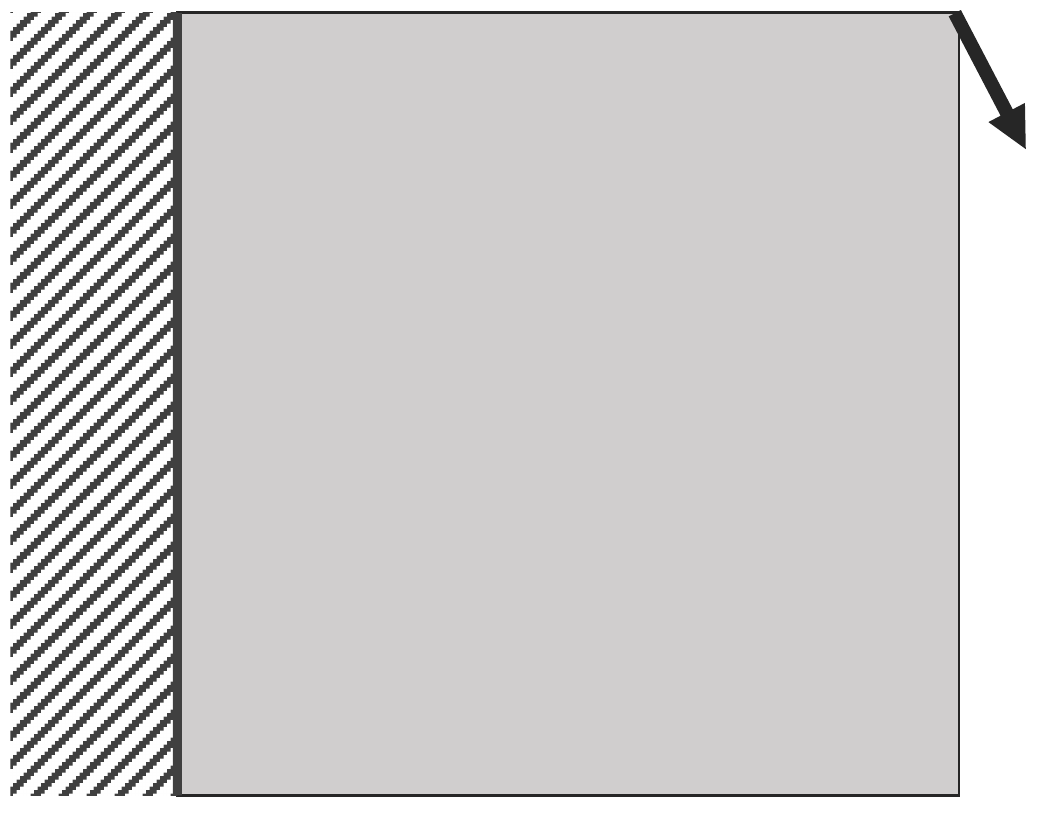}
    \caption{Problem boundary conditions considered for exemplifying the effect of {grey scale} and structural disconnections.}
    \label{fig:test1_BC}
\end{figure}

\begin{figure*}
   % \centering
    \begin{minipage}{0.245\linewidth}%
    \captionsetup{type=figure,justification=raggedright,singlelinecheck=false}
    \includegraphics[width=\linewidth]{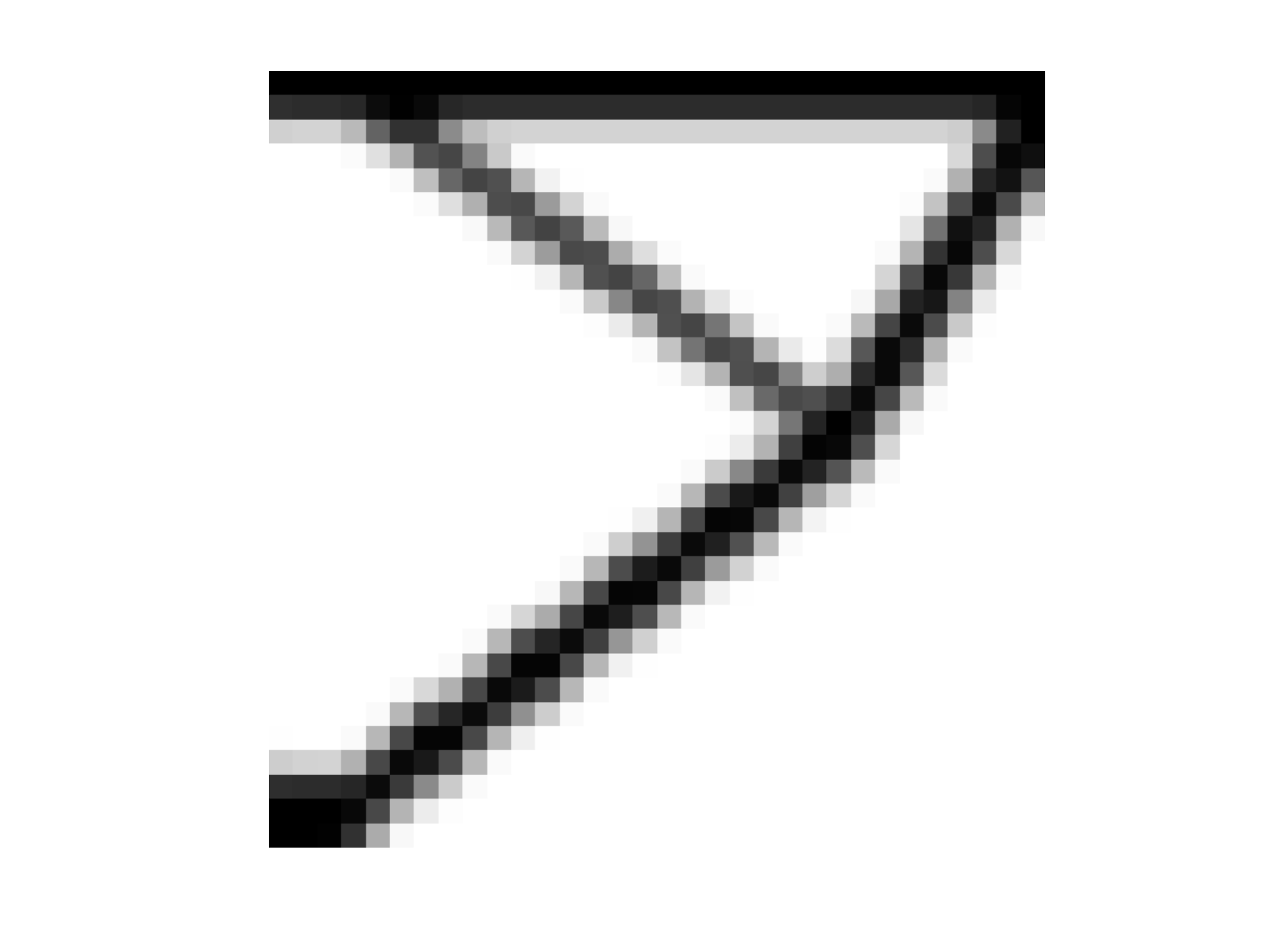}
    \subcaption{{Grey scale} connected \\ structure (compliance 90.04).}
    \end{minipage}
    \hfill
    \begin{minipage}{0.245\linewidth}%
    \captionsetup{justification=raggedright,singlelinecheck=false}
    \includegraphics[width=1.0\textwidth]{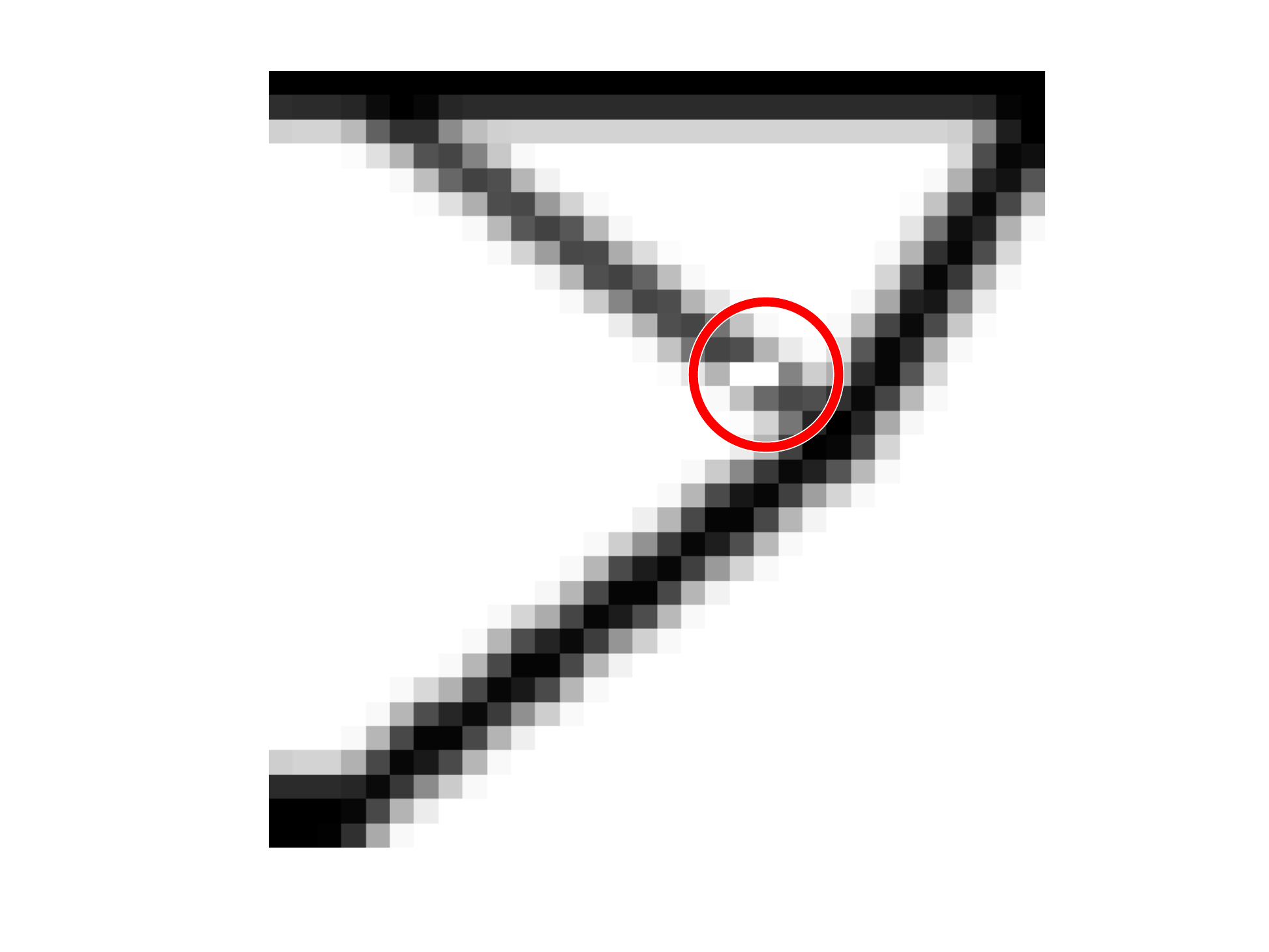}
    \subcaption{{Grey scale} disconnected structure (compliance 101.63).}
    \end{minipage}
    \hfill
    \begin{minipage}{0.245\linewidth}%
    \captionsetup{justification=raggedright,singlelinecheck=false}
    \includegraphics[width=1.\textwidth]{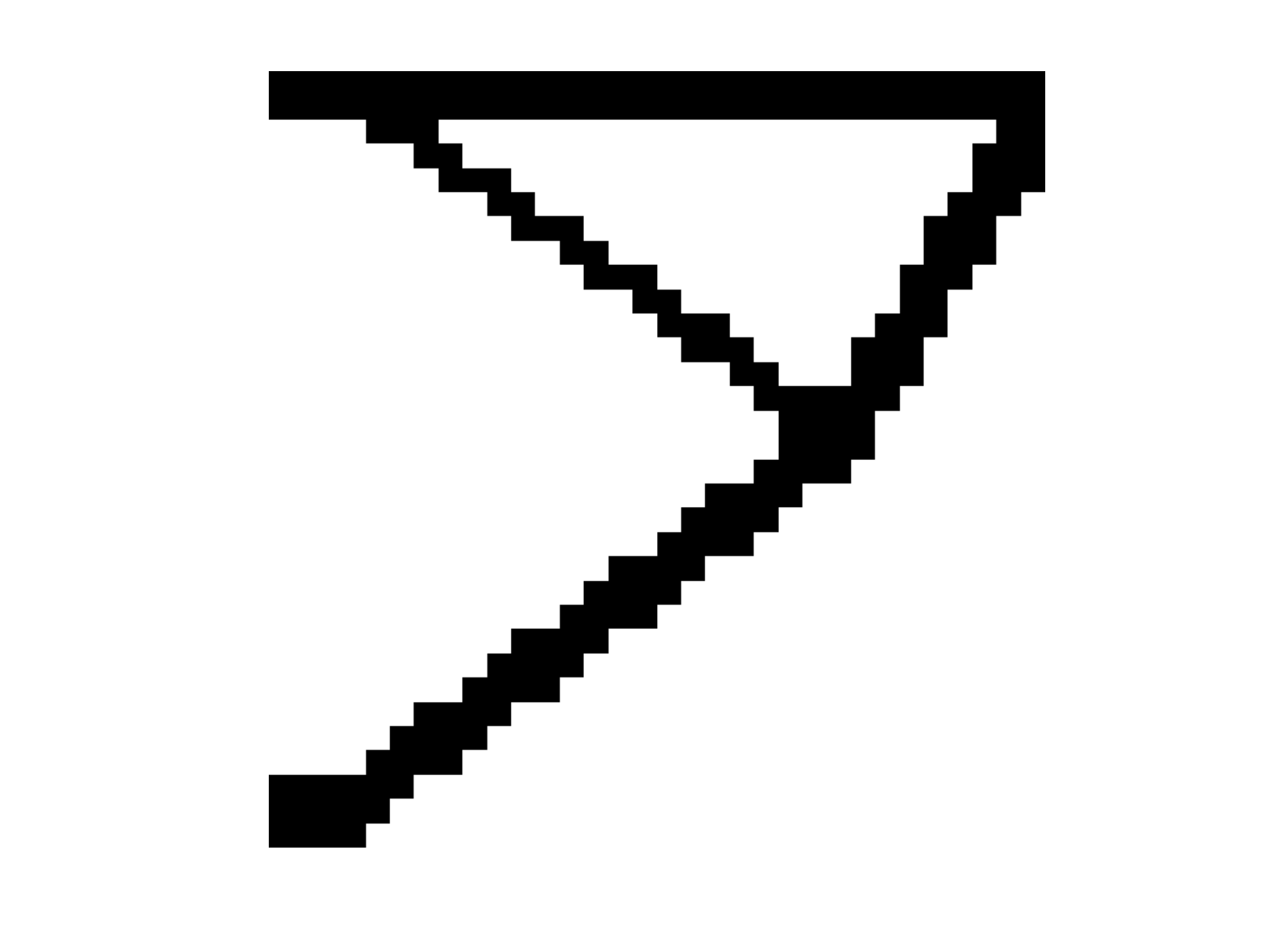}
    \subcaption{Black-white connected \\structure (compliance 66.16).}
    \end{minipage}
    \hfill
    \begin{minipage}{0.245\linewidth}%
    \captionsetup{justification=raggedright,singlelinecheck=false}
    \includegraphics[width=1.0\textwidth]{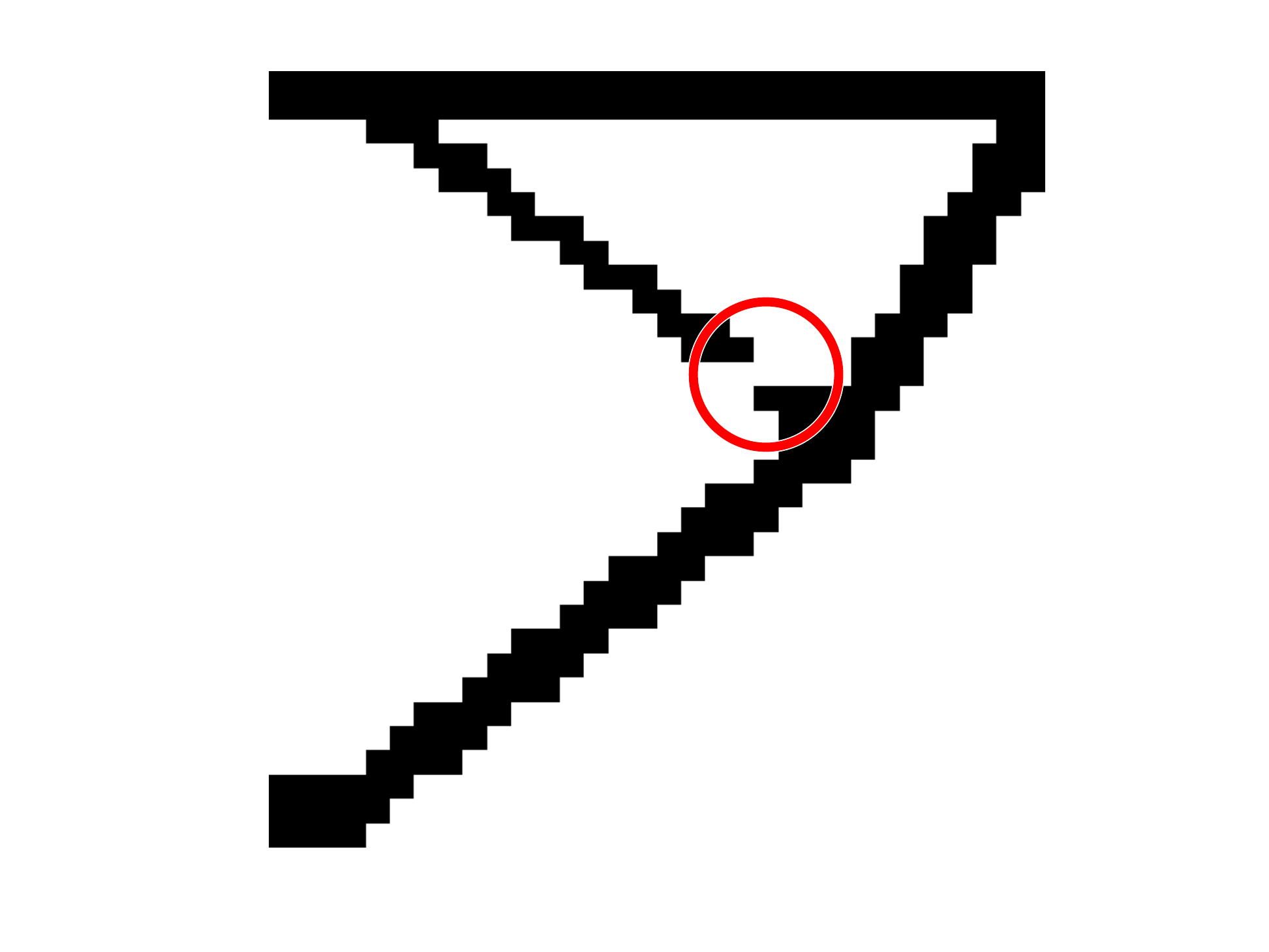}
    \subcaption{Black-white disconnected structure (compliance 199.46).}
    \end{minipage}
    \cprotect\caption{Compliance minimisation example for boundary conditions in Fig. \ref{fig:test1_BC} (subject to a {volume fraction} constraint of 0.2) illustrating the effect of disconnections on a 32x32 mesh. The {grey scale} structure (a) is obtained by \verb|top88(32,32,0.2,3,1.5,2)| (\citealt{Andreassenetal2011}). Disconnections are imposed (b) and thresholding is applied to obtain the 0-1 counterparts (c)-(d).}
    \label{fig:test1}
\end{figure*}

To illustrate both why such disconnections may occur and their effect on the structural performance, a simple test case inspired by the type of problems considered by \cite{YuHurJungJang2019} and \cite{NakamuraSuzuki2020}, is presented in Fig. \ref{fig:test1_BC}-\ref{fig:test1}. By reducing the density of two of the elements in the original structure (\ref{fig:test1}(a)), the central bar is almost disconnected completely (\ref{fig:test1}(b)). Applying volume-preserving thresholding (\citealt{SigmundMaute2013}) the corresponding solid-void structures (\ref{fig:test1}(c)) and (\ref{fig:test1}(d)) are obtained, where a full disconnection is now obtained. The structural compliance with respect to the boundary conditions (Fig. \ref{fig:test1_BC}) is indicated for each of the four presented structures. The presented test case is as such modelled on a square domain with a clamped left side, subjected to an external single-point load of horizontal magnitude 0.5 and vertical magnitude 1.0 applied to the top right node.

Table \ref{tab:test1} presents the mean average density error (MAE) as well as the relative increase in compliance (Gap) when comparing the connected and disconnected structures for both the {grey scale} and black-and-white designs. Firstly, it can be observed that for the {grey scale} structures, if the MAE was used to assess the difference between the two, they would be nearly identical. However, the compliance of the semi-disconnected structure is 12.9\% higher than for the originally connected. After thresholding to fully black-and-white designs this effect is intensified, as the MAE remains below 0.4\%, but the compliance is more than doubled. Therefore, if the model is trained with an increased focus on minimising errors related to image-reconstruction (e.g. MAE or Binary-Cross-Entropy) there is a risk of overlooking adverse effects when it comes to model-performance. The works of \cite{LuoZhouetal2021} and \cite{BehzadiIllies2021} corroborate this suspicion, as physical performance or topology awareness is included in the {loss function} of the direct design model and a reduction in structural disconnections is observed. \cite{Halleetal2020} further considered fully unsupervised learning for direct TO where no disconnections are observed in the illustrated examples, but occurences of discontinuities and a loss of fine features are still presented. Note that with increased physical information embedded in training, FEA is needed each time the {loss function} is evaluated, making the actual training procedure much more computationally expensive. This could, however, reduce the overall cost of training data generation by obtaining higher accuracy using fewer optimised structures for training. 

\begin{table}[ht]
    \centering
    \caption{The relative pixel-wise (MAE) and compliance error (Gap) between the original structure and the disconnected version for the {grey scale} and black-and-white cases from Fig. \ref{fig:test1}. }
    \begin{tabular}{c|cc}
    \toprule
         &  \textbf{{Grey scale}} & \textbf{Black-and-white}\\\midrule
        \textbf{MAE} & $0.0014$ & $0.0039$  \\
        \textbf{Gap} &0.1288& 2.0148\\\bottomrule
    \end{tabular}
    \label{tab:test1}
\end{table}

By presenting both the {grey scale} and 0-1 designs, other important aspects of comparing optimised structures are also highlighted. Firstly, considering the compliances of the two connected structures from Fig. \ref{fig:test1}, volume-preserving thresholding improves the structural performance significantly. Secondly, a partial disconnection by a low-density region in {grey scale} may result in a full disconnection when thresholded to a 0-1 design, which leads to a significant increase in compliance. As such, thresholding is important to reveal the true structural performance. {Therefore, in line with the recommendations of \cite{WangZhaoZhou2021}, solid-void designs are advised for a fair comparison of optimised structures.}\\

Based on the presented {example}, it is explained why MAE alone is not sufficient for either training a direct design model nor evaluating performance of a solution framework, as this measure may erroneously overestimate the performance of a structure. Further, the degree of {grey scale} may influence compliance comparisons between structures. Either a structure with more {grey scale} can be at a disadvantage or it can fail to capture a crucial structural disconnection. Hence, it is clearly a fundamental mistake to compare discrete to {grey scale} designs, or vice versa.\\

\cite{Bieleckietal2021} proposed an extended direct design approach utilising a {three-step} procedure. Given the problem-defining boundary conditions a DNN is first trained to realise a structure by determining the material distribution within the design domain. Thereafter a CNN is used for structure refinement achieving reduced {grey scale} and smoother boundaries. In the last step conventional TO is applied for a maximum of 5 iterations to post-process the structure to ensure physical consistency and volume-constraint coherence, i.e. removing disconnections and ensuring that constraints are satisfied. The model was trained separately for 2D and 3D problems with fixed {mesh sizes} of 80x80 and 20x20x20 elements respectively. Similar conditions for sampling of training data were utilised in both cases, where different volume fractions and supports or loads in {corner nodes} constituted the sampling space. To avoid rigid body motion, 3 out of 8 and 6 out of 24 {degrees of freedom (DOFs)} were fixed throughout for 2D and 3D cases, respectively. In total 614,304 samples were optimised and used for training for the 2D case, while for the more time-consuming 3D-case the training set size was restricted to 45,000 samples. Test-cases used for model-assessment are obtained by sampling from the same {problem space} as for the training data generation. Comparative results are not reported across all 1,000 test samples, but a significant speed-up is obtained and for the presented results the compliance values are at least as good as those obtained by conventional TO. This holds true for both the 2D and 3D cases, but as a significantly smaller fraction of the {problem space} is used for training the 3D-model, the overlap between training and test problems is expected to be smaller. Due to the large number of optimised structures needed for training, the construction of the model is computationally expensive. Further, as the problem instances considered are sampled within a very restricted subset of possibilities that utilise the same coarse {mesh resolution}, transferring this framework to problems outside the training-instance distribution is expected to increase the computational cost further.\\

{A common trend amongst the papers within the direct design category is the requirement of a large set of TO-optimised structures, which becomes computationally expensive to collect and is likely the reason why these papers only consider very coarse scale fixed meshes and few variations in boundary conditions. Even with a large amount of training data and a very restricted problem space most works present results with poor structural performance. The use of image-reconstruction type loss-functions only is a popular way of training the presented ANNs, significantly reducing the training time compared to if structural performance was to be integrated. Image-based errors do, however, not reflect the quality of a structure and thus the network learns based on an incorrect measure. There are other reasons for why the premise of this application category is flawed, which will be covered later in Section \ref{sec:2} and Section \ref{sec:666}.}

\subsubsection{Acceleration}
The application of AI-methods for accelerating TO is receiving increasing attention, where approaches both aim at limiting the number of iterations and complex computations needed within a conventional iterative optimisation procedure. The strategies in this category offer a more diverse profile than for the previously described direct design models, but there are some key similarities in the motivational ideas behind the presented works.

\paragraph{Sensitivity analysis}
An AI-method is considered to apply to sensitivity analysis when the aim is to replace or reduce the need for exact evaluations of sensitivities. Some of the works (\citealt{AuligOlhofer2013,AuligOlhofer2014,AuligOlhofer2015}) contained in this section are only applicable, in the sense that they actually facilitate a speed-up, in cases where the sensitivities are difficult to obtain by conventional FEA and adjoint analysis. The claim of existence of such TO problems is often heard, but seldom exemplified. Other works (\citealt{ChiZhangetal2021,QianYe2020,KesKirNara2021}) try to reduce the computational load of, or completely eliminate, FE-analysis needed in the TO process.\\

Common for these approaches is that they aim to train a model to approximate complex computations by some functional relation. With this purpose some feed-forward neural network is constructed and trained as a regression model using supervised learning. Typically the network inputs consist of at least the current element densities. In some cases the loading conditions are also included and for procedures where FEA is still present in some form, displacement or strain energies are also supplied.\\

A single pass through the network may apply to one individual element, a patch of elements within the structure or all elements in the structure simultaneously. Approaches only considering subsets of elements at a time can allow for increased generalisation ability in that mesh- and problem-dependencies are potentially reduced, but on the other hand, important global information may be overlooked. Further, FE-analysis of a single structure {and its density histories} can provide a larger set of training samples with varying characteristics. As such, {data generation} is in most cases significantly cheaper than for the direct design models and there is potential to naturally capture greater input-diversity. By considering such sub-structures, the similarities between training and test data could be expected to increase, even with vastly different boundary conditions and load cases.\\

 \cite{Lee_etal2020} proposed a solution framework based on the conventional {Optimality Criteria (OC) method}, where two separate CNN-models are trained to predict compliance and volume fraction respectively. For compliance this means that the need for FEA to evaluate the structural integrity is eliminated, replacing the computations with a less complex functional approximation. The conventional computation of the volume fraction is of linear complexity which is now replaced by some non-linear function represented by the corresponding CNN. The overall idea is that these neural networks will reduce the computational load of each iteration in the optimisation process, resulting in a significant speed-up. Most of the presented experimental  results are focused on the network's ability to predict volume fraction and compliance for a given structure, and thus the integration of the model in a TO process, where element sensitivities are needed, is not detailed. The MBB-beam and cantilever beam with fixed {mesh discretisation} and varying {volume fractions} are considered throughout the paper, both for training and testing. A full assessment of the method performance is therefore difficult, due to large similarities between training and test cases. \cite{Papadrakakis1998} and \cite{Sasaki2019} similarly presented ANNs trained to predict objective and constraint values of a structure, aimed at replacing the fitness-evaluations in each iteration of a GA framework. \cite{QiuDuYang2021} trained networks to iteratively remove material from a fully solid domain, similarly to evolutionary structural optimisation, but without the use of FEA in the actual optimisation procedure.\\

\cite{AuligOlhofer2013}, \cite{AuligOlhofer2014} and \cite{AuligOlhofer2015} focused on designing regression-type ML-models for predicting sensitivities when the adjoint approach is unattainable (not exemplified in their work) such that finite differencing is the only alternative. Standard compliance minimisation problems are used as examples, for which the formulas for exact sensitivities are known. The model inputs are related to the element densities and displacements (computed by FEA). In compliance minimisation the exact gradient of an element with respect to the objective is a function of these same features. This means that a good performance in terms of {sensitivity accuracy} can be expected for this exact problem formulation, but no conclusions about transferrability to other formulations can be made. A framework for further limiting the computational cost of the finite differencing alternative is also presented. Here the computational load of FEA is reduced by adaptive sampling of elements needed for exact evaluation, reducing the degrees of freedom in the FEA. This strategy is, however, not related to ML {or NNs}, and thus outside the scope of this review.\\

If the desired effect of the model is to reduce the computational load of FEA, but not necessarily completely removing this analysis technique from the optimisation procedure, another option for training the model is to facilitate for online learning. Online learning refers to when the model is not trained on pre-collected data, but rather trained during application to adapt to a specific problem. For the considered purpose, this entails that the model is trained during the optimisation run where FEA is then only completed in a subset of iterations and the obtained solutions are used in a sequential transfer learning procedure with an increasing number of {data samples}. \cite{ChiZhangetal2021}, and similarly \cite{ZhangChiPaulinoetal2021}, presented such an approach where a transfer learning based procedure is conducted after each set of new training data additions as to iteratively make the model more precise. The authors propose to perform the optimisation as a two-scale approach where a coarse grid version of the structure is subject to FEA at each iteration, while the trained model is applied to map these results to a finer mesh where FEA is only applied at a subset of the iterations. {These approaches are as such not only concerned with sensitivity analysis, but utilise a multi-level TO approach to obtain the computational reductions associated with the sensitivity analysis.} The presented results show promise and accuracy of the fine grid sensitivity predictions appear to improve with this online-approach. The approach and resulting speed-up are similar to those of multi-resolution techniques (\citealt{Groen16,Nguyenetal2012}).\\

{The main challenges for ANNs aimed at reducing the computational cost spent on FEA during optimisation are two-fold. Firstly, some approaches build on erroneous premises where the conventional alternatives are assumed less efficient than they really are, leaving these approaches redundant. Secondly, models developed for very specific problem-instances become too restricted to be used as a general framework for TO. As for the multi-resolution online-learning approaches, these may be affected by coarse mesh restrictions of fine features when projected to a higher resolution. This phenomenon is common for the two-scale methods reviewed and will be covered in more detail when covering the upscaling category in Section \ref{sec:post-process}}.

\paragraph{Convergence}
If a model is trained to map between intermediate solution structures with the aim of reducing the total number of iterations in the optimisation procedure, it is said to pertain to accelerating convergence. Typical choices of methods are based on either a direct design-like model (\citealt{YeLietal2021,JooYuJang2021}) or some time-series inspired forecasting (\citealt{Kallioras2020}). The direct design-type models map the input {grey scale} image of an intermediate design to an almost converged structure. Alternatively, the time-series inspired methods consider the trajectories of the densities of the individual elements and seek to directly map each element from a given iteration to a close to converged state.\\

The first approach inherits most of the challenges associated with direct design models. If the full image is to be mapped at once, the constructed model is likely to be mesh-dependent. Ensuring diversity and accuracy for different problem characteristics is difficult and comes with a large computational cost related to {data sample} generation. It does, however, benefit from the fact that more descriptive data is available as network inputs. From the performed optimisation iterations, used to reach the intermediate structure, both the displacement-field and density-history of the elements are known and can be used as inputs to the model.\\

\cite{SosnovikOseledets2017} trained a CNN to translate the {grey scale} image of the structure obtained after $k$ {SIMP iterations} to a final black-and-white design. The training dataset was generated by running 100 iterations of SIMP on pseudo-random {problem formulations} on a 40x40 mesh. As inputs to the network the element densities at iteration $k\leq 100$ and the latest change in densities from iteration $k{-}1$ to iteration $k$ are supplied as two {grey scale} images. Different strategies for sampling $k$ in each training sample were tested, and the output target considered was the black-and-white image obtained by thresholding the optimised structure at $k=100$. The trained model is shown to outperform standard thresholding for the training dataset, but when tested on new problem formulations, heat conduction problems, the performances become similar. Performances are measured by binary similarities between structures, and no measures for compliance or volume fraction of the obtained structures are reported. Structural results are also illustrated for problems similar to the training datasamples on finer grids (up to 72x108 elements), but the effect seems to simply be smoothing of boundaries compared to coarser structures. \cite{JooYuJang2021} proposed a similar approach, but instead of mapping the full structure at once, their model divided the structural image into overlapping sub-modules, which then separately are mapped to an optimised sub-structure, and the complete structure is subsequently obtained by integrating over these sub-modules.\\

One benefit of considering the structural image as patches or a whole instead of element-wise is that some information about the interaction between the elements can be retained and learned by the CNN. In the time-series approach however, there is an assumption of independent density-trajectories of each element which might be problematic exactly because of the fact that the elements must have appropriate interaction to form a viable structure. Some of these interactions may be observable for the network given similarities in their iterative density-histories before the mapping is applied. A benefit of this approach is that the generalisation ability of the proposed method is likely to increase as there is less mesh- and problem-dependency reflected in the training samples. Success of this method does, however, depend on the expected iterative trajectories being similar even for different {problem definitions}. A moving structural member will lead to bell-type trajectories, meaning the direction of the density-trajectory of the concerned elements change several times during the full iteration history. The assumption is therefore that an element's density-trajectory during the first few iterations is sufficient to distinguish the elements for which this happens, no matter what structural problem is considered. \\

\cite{Kallioras2020} proposed a time-series approach where the iterative element density histories over the first 36 {SIMP iterations} were used as input to a neural network which individually maps the element densities to close-to-converged values to form a structure from which SIMP is continued until convergence. The model used is a Deep-Belief Network (DBN) which is a type of ANN where feature detection to achieve dimensionality reduction is conducted in each layer. As such, the input-vector of the iterative density history of an element is gradually reduced to a final density value throughout the network. The network was trained on {data samples} consisting of the iterative history obtained from solving versions of the cantilever and simply supported beams with different length-scales and discretisations. The number of finite elements in these training samples ranged from 1,000-100,000, where four sets of boundary conditions for two different length-scales were solved for each resolution. {When testing the model on problems different from the training cases, a computational speed-up is achieved. The speed-up is reported in terms of the number of SIMP-iterations needed to reach convergence, compared to the conventional approach. The obtained solutions have compliance values approximately matching those of the SIMP-obtained benchmarks.} It should be noted that the comparisons to the conventional SIMP-approach is done in {grey scale}, which as shown in Fig. \ref{fig:test1} may significantly underestimate the actual stiffness of a structure and thus the comparative results may not be representative.\\

{Common for the presented convergence applications is that they assume part of the iterative density-history early in the optimisation procedure is sufficient to determine the nature of the final result. This can be a problematic assumption for problem instances where structural members move during optimisation causing large changes in both the individual element densities and the density field as a whole. The time-series approach based on individual pixel density histories is especially unlikely to succeed, but there may still be potential for approaches considering the global design change (\citealt{MunozChinesta2022}). It is reasonable to assume that given previous iteration history it is possible to predict the density-change after a subset of consecutive iterations, but by eliminating significant parts of the iterative search one is likely to face some of the same challenges as for the direct design applications.}

\subsubsection{Post-processing}\label{sec:post-process}
AI-methods are considered as post-processing procedures when an optimised structure is used to generate the model input. As such, this application category pertains to methods for interpolating the given structure to a finer {mesh resolution} or shape optimisation and feature extraction for manufacturability purposes.

\paragraph{Shape optimisation}
When the aim of the formulated model is to alter the features of the obtained structure to ensure practical and cost-efficient manufacturability requirements are satisfied the method is said to perform post-processing by shape optimisation. Design aesthetics may also motivate such applications, where for instance \cite{Vulimirietal2021} considered TO for minimal compliance while adhering to some reference design for structural patterns like circles or spider-webs. \\

Two of the works presented in this category, \cite{LinLin2005} and \cite{Yildizetal2003}, each proposed versions of ANN designed to perform hole-classification in a TO optimised structure. \cite{LinLin2005} proposed a two-stage procedure where the first ANN is trained to recognise the underlying basic geometric shape of the hole, when given an input in the form of invariant moments describing the geometric characteristics of the hole in the optimised structure. The second stage consists of several ANNs, one for each basic geometry group defined for the first ANN, and they are each trained to fit a detailed shape template for the hole, within their basic geometry group. The input format used is a set of distance and area-ratios of the hole image represented in a dimension-independent manner, and the network uses this information to map the hole to one of twelve predefined geometric shape templates within the considered shape-category. \cite{Yildizetal2003} trained a single ANN which uses the {grey scale} image of the TO optimised structure as input and computes confidence measures for each hole in the structure, based on the perceived similarities to four basic feature templates. The identified hole-shapes are then used to formulate a feature based part model which is subjected to {shape optimisation} to obtain the desired final structural layout. The main challenges associated with the presented works relate to the fact that they both rely on manually defined sets of possible shapes. Further, as the sizes of these sets are limited it is unclear whether anything is gained from applying AI-based models to perform the identification tasks. \cite{Yildizetal2003} also found, by testing different ANN architectures, that the best results were obtained when the ANN only contained one hidden layer. This raises the question of whether alternative and simpler deterministic methods could achieve similar results.\\

\cite{Hertleinetal2021} proposed a direct design type GAN-model with integrated manufacturing constraints for additive manufacturing and paired it with a post-processing procedure utilising conventional TO. The inputs to the model consist of channels relating directly to the 64x64 mesh considered, indicating supports and loads as well as build-plate orientation to account for the manufacturability. The input encoding is constructed such that existence of material is encouraged in elements where the optimised structure is expected to have material, here defined by the locations of loads and build plate. The output from the GAN is a {grey scale} image representing the optimised topology. Training data {are} obtained running conventional TO (\citealt{Andreassenetal2011}) with an integrated overhang filter as presented by \cite{Langelaar2017}. It is also suggested that the resulting structure is post-processed by running some number of iterations of this conventional alternative, to further eliminate overhanging features and correct any compliance-related inaccuracies.\\

{Overall, this type of application of AI-technology is not well-studied in the literature. This might be due to the introduction of filters and manufacturability constraint in the TO problem formulation and solution process reducing the need for post-processing, or that alternative conventional methods for post-processing with satisfactory performance exist.} It would be of great benefit if ML could be used to extract CAD geometries from optimised designs, as many manufacturing methods require a format for structural representation which is not directly attainable from density-based designs. The viability of obtaining such a model is, however, not guaranteed. {There is a body of literature on reverse engineering methods (\citealt{Buonamici2018}). These methods reconstruct CAD models from acquired 3D data in the form of triangle meshes or point clouds. Some methods produce constructive solid geometry (CSG) models (\citealt{Duetal2018}). These methods are often based on detecting primitive shapes in the input (\citealt{LiWuetal2011}) whereas \cite{EckHoppe1996} generated B-Spline patches from the input. Such methods could be exploited for post-processing of TO optimised structures.}

\paragraph{Upscaling}
The works belonging to the post-processing upscaling category typically consider a coarse grid structure optimised using conventional methods and apply some type of neural network to translate this structure to a finer mesh. There are various approaches to how to format the input to the considered model where \cite{Wangetal2020} and \cite{YuHurJungJang2019} evaluated the entire structural image of element-densities as input while \cite{Napieretal2020} and \cite{Xueetal2021} divided the structure into patches of element-densities that are processed individually but may have some overlap in terms of what elements belong to each patch. The latter option is likely to be the most beneficial in terms of generalisation ability, especially as the models have the potential to be more or less mesh-independent. Further, the patch-based approach may allow for fewer optimised high-resolution structures to be used for training data, and thus overall significant computational cost-savings may be achieved. One concern is, however, that the applied model does not have any concept of the structure as a whole, such that only boundary fine-tuning and no topology refinement is obtained.\\

\cite{KalliorasLagaros2020} proposed a method (DL-scale) that somewhat differs from this approach as they apply deterministic upscaling, iteratively paired with the DBN convergence acceleration framework proposed in \cite{Kallioras2020}. This work does therefore not apply AI-methodology for the actual upscaling, but is mentioned here as the reported results still reflect some of the common challenges within this category. Even though it is true that they observe significant speed-up for increasingly finer grids it becomes evident that the solutions obtained by the modified approach has a reduced capability of capturing finer features, when compared to the corresponding SIMP optimised structures. Further, for several of the reported cases, the level of {grey scale} appears to be higher for the DL-scale obtained structures, and as the minimum compliances compared in each test case are computed for {grey scale} images, it is unclear whether the overall better objective values obtained using DL-scale actually are representative. The lack of finer structural components when applying upscaling is a common occurrence in the literature, and works like \cite{Wangetal2020} illustrate how the attempt to capture fine features might lead to structures with a large degree of blurry {grey scale} areas or even structural disconnections. Essentially, details on the fine scale are limited by the coarse scale resolution, which effectivly works as a crude length-scale constraint.\\

\cite{Elingaard2021} proposed a CNN for mapping a set of lamination parameters on a coarse mesh to a fine scale design promoting very fine features. The network is as such used as a computationally efficient substitute for de-homogenisation (\citealt{PantzTrabelsi2008,Groen18}) to overcome the current bottleneck in extraction of fine scale results in homogenisation-based topology optimisation. As inputs to the network the orientations from a homogenisation-based TO solution are used. The network is then used to upsample this information to an intermediate density field, which is post-processed using a sequence of graphics-based steps running in linear time to obtain the final high-resolution one-scale design. Unsupervised training is utilised to avoid the need for generating expensive targets and cheap input data generation for training is ensured by sampling from a surrogate field of low-frequency sines. The training is as such performed independently of the physical properties of the underlying structural optimisation problem, which makes the method mesh and problem independent. By numerical experiments, it is found that this approach achieves a speed-up of factor 5 to 10, compared to current state-of-the-art de-homogenisation approaches.\\

{Common for the presented two-scale approaches aimed at translating structural information from a coarse to a fine grid, using the same measures, mainly densities or sensitivities, is that the coarse scale mesh imposes length-scale constraints on the fine grid. This means that little information is gained by utilising ANNs to perform this mapping, when compared to conventional interpolation techniques. This challenge does not occur if the ANN is applied for de-homogenisation, as here it serves as a tool for replacing a computational process which is a part of a pre-existing upscaling scheme where details are constructed from coarse scale information based on predefined rules.}

\subsubsection{Reduction}
The typical approach for achieving reduction or problem re-parameterisation by use of AI-methods is to construct one or more inter-connected neural networks with the aim of representing a structure using fewer design variables and thus decrease the computational load of the optimisation procedure. This can be done by training a VAE for feature extraction and exploiting the reduced dimensionality of the obtained latent space to conduct the optimisation on this latent vector. Alternatively, the network can be constructed as a direct surrogate for the optimisation process such that the training of the network is equivalent to solving the given optimisation problem exploiting that the parameters and biases of the network are sufficient as design variables. \\

\cite{Guoetal2018} considered a multi-objective thermal conduction problem for which a VAE is trained in a supervised manner, with the aim of {minimising} the reconstruction-error of the encoder-decoder network. The model is then tested by integration in various conventional optimisation frameworks, including gradient-based methods, genetic algorithms and hybrid versions of the two. By encoding the intermediate structure, design-updates can be executed in the reduced latent space. The new latent vector can then be translated to an interpretable structure by the decoder, which is next subjected to physical analysis. As such, FEA is still needed for the full design space, using the same {mesh discretisation}, meaning the computational cost of computing objective and sensitivities remains the same as for the conventional methods. Nevertheless, there might be a potential gain in performance by reducing the number of iterations required, as the number of FEAs reported to reach convergence varies between the different solution frameworks tested. However, few test cases are reported and little comparison to state-of-the art procedures is conducted.\\

\cite{ChandrasekharSuresh2020} used an ANN to re-parameterise the density function, and thus in principle making the density representation independent of the FE-mesh. When integrating the new structural descriptor into a conventional solution framework, the weights and biases of the ANN become the design-variables that are optimised through unsupervised learning with a {loss function} corresponding to a weighted sum of structural compliance and volume-constraint violation. {This is equivalent to conventionally optimising a new design representation, meaning that the network is not subjected to any actual learning. Thus, this is an example of using an ANN without the learning aspect.}\\

Later, \cite{ChandraSuresh2021} showed how this framework can be extended to a multi-material TO problem where the distribution of two or more materials within the structure is obtained simultaneously with the optimised topology, and \cite{ChandraSuresh2022} added a Fourier-series extension to the ANN to impose length-scale control. In either case, FEA is evaluated on the same FE-mesh, which in each iteration is constructed by sampling densities for the needed spatial coordinates using the ANN. As such, conventional physical analysis on a discretised grid is still necessary to compute the sensitivities of the objective (in this case also the {loss function}) with respect to element densities, after which the sensitivities with respect to network parameters can be determined by classical back-propagation. A promising feature of this application is that because of the analytical density-field representation, sharper structural boundaries can be obtained. Currently, however, the structures are projected on a fixed FE-mesh for analysis which means that the boundary effectively is blurred. Further, there is a loss of {fine features} in the structures obtained by the new solution procedure, and detail does not appear to increase much with finer meshes. The results presented in \cite{ChandraSuresh2022} also indicate that artefacts from the coarse discretisation may cause non-physical structures in the upscaled results. Moreover, as $>$90\% of the optimisation time is spent on the FEA, this approach is unlikely to provide any promising speed-up unless additional measures are implemented to reduce the efforts needed to complete this evaluation process. \\

\cite{DenTo2020} presented a re-parameterisation approach similar to that of \cite{ChandrasekharSuresh2020}, but with an increased focus on enabling representation of detailed 3D-geometries. Their method is coined \emph{deep representation learning} and several different test-cases illustrate the increased ability to achieve structures including finer features. A comparative study to conventional TO is not detailed in the article, but the results do encourage further exploration of this method's capabilities. Additionally, applications related to post-processing, and more specifically extraction of CAD models for manufacturability, could potentially benefit from this approach.\\

Other similar versions of re-parameterisation applications are also attempted in the literature. \cite{ChenShen2021} perform online training of a GAN to obtain an optimised structure. The model is trained to firstly ensure volume-constraint satisfaction and secondly compliance value minimisation in an iterative manner. \cite{Hoyeretal2019} and \cite{ZhangZhao2021} altered the approach to directly enforce the constraints in each iteration, reducing the {loss function} to compliance only. \cite{DengTo2021} replaced the level-set function with an ANN, \cite{Zehnderetal2021} combined the method with a second ANN aimed at predicting displacements to achieve mesh-free TO, and \cite{Greminger2020} ensured manufacturability in each iteration by manipulations in the latent space of a trained GAN. \cite{HayashiOhsaki2020} and \cite{ZhuGuoetal2021} performed reparameterisation by reinforcement learning for truss structure optimisation. As most of these approaches still perform FE-analysis on the full mesh in each iteration, reported speed-ups are mainly caused by a reduction in the number of iterations until convergence. Another common trend for these works is that the resulting structures have fewer fine scale features than the corresponding solutions obtained by conventional TO. One could therefore speculate whether the reduction in iterations is a result of the re-parameterisation causing a perceived larger {filter radius} or coarser mesh.\\

{Most of the works utilising ANNs for reductions in the dimensionality of design representation do not rely on typical learning techniques, as the network is re-initialised before optimisation each time. Here the NN architectures are simply used as a reparameterisation of the density field which is then subjected to a conventional optimisation procedure. An evident challenge for such approaches is the decreased ability in representing fine features when using fewer network-defining parameters.}

\subsubsection{Design-Diversity}

Generative design is the process of exploring different design options satisfying structural performance requirements and selecting a suitable subset fulfilling various specifications. A good subset of structures would present visually different good-quality design candidates, providing the option of selecting the final design based on other practical or visual demands not integrated in the optimisation model. Such demands could be the personal preference of a designer wanting a visually pleasing structure, which may not be directly quantifiable by mathematical constraints. To this end, ML-applications for design-diversity are aimed at maximising the aesthetic variety of the search space in the exploration phase or at determining the best subset in the selection phase.\\

\cite{RawatShen2018}, \cite{RawatShen2019}, \cite{ShenChen2019} and \cite{RawShe2019} presented a series of papers considering the same GAN-CNN paired framework. The GAN is here trained using 3,024 conventionally optimised structures to generate new unseen structural variations, while the CNN is trained to predict the volume fraction, penalty parameter and filter radius corresponding to these new solutions. The same fixed boundary conditions are considered throughout the papers, but only the 2D formulation is considered in the first three while \cite{RawShe2019} extend the model to 3D. The proposed framework provides a way of exploring the parametric solution space for a single problem requiring fewer direct optimisations, thus reducing computational time. In this manner, a larger number of design-options for a structure can be investigated. The obtained structural results for the CNN-GAN pairing similarly to the direct design models exhibit some disconnections and noisy boundaries, motivating a post processing procedure exploiting different filters to obtain a smoother design. This post-processing procedure is utilised throughout all above mentioned papers, and is found to significantly improve the generated designs, but not completely remove all occurrences of noise or disconnected features.\\

\cite{OhJungKim2019} similarly integrated a GAN in the exploration process to more efficiently generate new and different designs by replacing some of the topology optimisation runs needed. The network is trained to generate viable wheel designs that appear different from some given reference structure, such that it can expand the set of diverse designs more quickly. \cite{SunMa2020} and \cite{JangYooKang2020} proposed alternative processes for exploration in generative design utilising reinforcement learning to maximise the design diversity. \cite{SunMa2020} employed different exploration tactics to alter the search trajectory of density-based TO methods by integrating reinforcement learning in the TO-process. \cite{JangYooKang2020} combined reinforcement learning for parameter selection with GAN, similarly to \cite{OhJungKim2019}, for faster generation of new designs. \cite{Yooetal2021} expanded on the generative model from \cite{OhJungKim2019} where {ANNs are} also applied for both upscaling the 2D design to a 3D CAD design and prediction of physical performance. \\

A natural extension to design diversity is to consider multi-objective optimisation problems, where the exploration and selection relates to the determination or selection of a variety of options from the Pareto-front. \cite{Satoetal2019} utilised clustering and association rule analysis for selection of a beneficial subset of structures. The fundamental idea is to train a machine learning model to recognise determining similarities and differences between structural composition and performance, such that comparable designs can be grouped together. The selection of a limited subset which contains designs spanning a wide variety of structural options can then be obtained by sampling from each of the obtained groups. \\

{ANNs for design diversity appear to have value in creating visually different designs. Much like for the direct design applications, the structural integrity of the newly generated designs cannot necessarily be trusted, and thus post-processing would be advised before the final design selection.}

\subsubsection{Other applications and physics}
There is a wide variety of different {NN}-applications in the literature, and not all structural TO frameworks were deemed to fit within the frames of the presented categorisation. Some of these are, however, still worth mentioning as they contribute to a more complete picture of the current state of the field.\\

Even before the real emergence of ML-assisted TO as its own field, a few preliminary works considered using NN-like models to support size and shape optimisation. \cite{AdeliPark1995}, \cite{ParkAdeli1995} and \cite{AdeliPark1995b} presented one of the earliest works utilising {NN}-models for structural optimisation. A neural dynamics model was presented, corresponding to an ANN with one variable layer and one constraint layer, meaning that the network size is related to number of design variables and constraints in the optimisation problem. \cite{Papadrakakis1998} and \cite{PapaLaga2002} later proposed a NN to replace the structural analysis within an optimisation framework based on Evolution Strategies (ES), obtaining a non-gradient optimisation procedure. The approach proved to provide significant speed-up compared to a ``standard'' ES optimisation algorithm, a family of methods later judged insufficient (\citealt{SigmundGA2011}). {\cite{Luoetal2020} proposed another non-gradient TO framework, the Kriging-based MFSE method, utilising Gaussian process regression to build a surrogate model and a material-field series expansion representation of the structural design. Results indicated that a much larger number of FE-evaluations are needed to obtain convergence compared to gradient-based TO methods.}\\

{\cite{Lynchetal2019} and \cite{Jiangetal2020} proposed ML-strategies to aid in tuning of parameters used in TO by SIMP and MMC to limit the number of re-optimisations needed when uncertainties in the appropriate choice of optimisation parameters is present.} \cite{PerKesWhi2020} tested different clustering and sampling approaches used for subset-selection within a visualisation framework aimed at illustrating the solution space for TO problems and the relationship between changes in boundary conditions and optimal solutions. \cite{NieJiangKara2020} presented a CNN to predict stress-field distribution of a cantilever structure with external loads applied to the free end of varying magnitude and orientation and a selection of domain shapes (rectangular, trapezoid and holes). The isotropic material properties, {mesh discretisation} and supports were considered fixed. 100,000 instances were used for training, each requiring FEA to obtain the target values. As the test cases presented were sampled from the same restricted {problem space} as the training data, no efforts were exhibited to ensure clear distinction between training and test data. This means that accuracy in the obtained predictions does not necessarily prove the model has learned anything.\\

\cite{LiShietal2021} presented a GAN utilising online training to replace stochastic alternatives in failure sampling during {subset simulation} for optimisation of periodic structures. \cite{YimLeeKim2021} proposed an ANN for predicting topology and end-effector location of a planar linkage mechanism given the path description. \cite{Barmadaetal2021} utilised a VAE to predict magnetic field distribution of a die press to accelerate optimisation of a die press with an electromagnet for orientation of magnetic powder. \cite{Bonfatietal2020} proposed a CNN for predicting the deformation properties of an image of mechanical actuators within a Monte Carlo/simulated annealing strategy for optimisation. As such, this approach is effectively an acceleration framework, but due to the image-related nature of the approach and the large number of training samples, nearly one million structural images, many of the same arguments against claims of improved performance compared to conventional TO methods as for direct design approaches, like e.g. \cite{Bieleckietal2021}, can be made. \\

{NN approaches have also become popular and largely adopted by researchers outside the structural optimisation field. Probably, due to the lack of knowledge or insight into TO, partly due to simple access to NN and GA schemes and partly due to reviewers that are not aware of TO advances, there is a rapidly growing trend for such papers in physics oriented journals. Examples are \cite{AbueiddaAlsmarietal2019}, \cite{ChenGu2020}, \cite{GuChenBuehler2018}, \cite{KimYang2021}, that solve typical linear TO problems on coarse grids for crack-propagation and \cite{JiengFan2019_nano1}, \cite{JiangChenFan2021_nano2}, that solve nano-photonic grating problems.}\\

{\cite{AbueiddaAlsmarietal2019} and \cite{GuChenBuehler2018} trained a CNN to predict the mechanical properties of a two-phase (soft or stiff material) chequerboard composite material to replace the need for FEA within a GA optimisation framework. \cite{KimYang2021} proposed an extended framework incorporating active learning such that the predictor can adapt to the specific problem considered, during optimisation. The additional data samples used in this iterative transfer learning are obtained by validating the selected {solution pool}, obtained by convergence of GA using the DNN for function evaluations, by FEA. The true target values for these, ideally well-performing, composites are then known and can thus be used to fine-tune the DNN before the next GA run.\\
\cite{ChenGu2020} proposed a general-purpose inverse design approach utilising a predictor-designer DNN-pairing. The predictor is trained to approximate a physics-based model or complex function and the designer utilises this learned mapping to perform optimisation of some specified desired properties. An integrated feedback-loop allows for continued improvement of the predictor as a response to the output from the designer. Maximisation of toughness in a 2-phase composite material subjected to individual base material volume constraints was presented as a case study of the presented framework. A fixed discretisation of 16x16 elements is considered, and for three different volume fractions one million composite designs are sampled and their toughness evaluated by FEA to form the dataset. 800,000 of the samples for each volume fraction are used for training and 200,000 for testing. From the active learning from the feedback loop mechanism, new training and test samples of higher value toughness are evaluated by FEA and added to the dataset during optimisation.}

\paragraph{{Inverse homogenisation} and composites}
Inverse homogenisation (\citealt{Sigmund1994}) {and} design of metamaterials is another rapidly growing application area of AI in TO – not only in structural applications but also e.g. in optics and nano-photonics (\citealt{JiengFan2019_nano1,JiangChenFan2021_nano2}). A feature of {inverse homogenisation and} meta material design problems that may make them better suited for AI approaches compared to structural problems is the limited number and position independent nature of load cases for such problems. Typically, in order to e.g. determine effective mechanical properties of a periodic material, just three load cases are needed in 2D and six in 3D, which in each case are independent of the design and geometry of the unit cells. Hence, the need for training data is significantly reduced. At the same time, however, variability in the outcome is also significantly reduced raising the question of whether an AI-approach is even needed for such problems. For example, there is no need for complex training if the goal is to provide stiffest possible microstructures for given macroscopic stress fields. In this case, analytically optimal multi-scale microstructures, rank-N laminates, are known and can be converted to simpler single-scale microstructures with little effort and loss as described in e.g. \cite{Traffetal2019}.\\

Similarly to direct design AI-applications for structural TO, \cite{Kollmannetal2020} trained a CNN to achieve iteration-free TO by predicting the optimal material layout directly from given problem-defining parameters for {grey scale} microstructure design problems. \cite{WangChanAhmedetal2020} developed a VAE-ANN pairing to transform the inverse design problem for unit cell solid-void microstructures to sequences of simple vector operations in the latent space. The VAE is for this purpose trained to achieve a smooth latent space capable of representing geometric information about different microstructures.  \cite{SuiGuoetal2021} employed reinforcement learning to automate the {design process} of digital materials and \cite{Garlandtetal2021} used a CNN to predict properties of solid-void lattice {materials}, both to be incorporated within non-gradient GA frameworks.\\

As the microscale design problem offers reduced {design freedom} compared to structural TO, utilising a large number of training samples to train a direct design-like ML-application increases the likelihood of training and test-data overlap. Therefore, the measured performance of such ANN-frameworks may be questioned. The ability to extract geometrical families as in \cite{WangChanAhmedetal2020} is, however, a beneficial trait of the presented approach. The works utilising ANNs to remove the need for FEA within a GA framework claim success based on the NN's ability to outperform conventional GA for problems with very few design variables, ignoring the large number of training samples used and the computational effort this entails. Utilising several thousands of samples to train a NN to speed up the optimisation of a 7x7 element mesh by claiming negligible training time gives a disproportionate representation of actual gain as a large part of the full solution space will be covered already in the training set. As such, the presented results do not actually prove the claimed viability of such GA frameworks. Further, these observations tie into how these works are found to suffer from similar issues to those discussed for other direct design approaches. \\

\cite{GuoYang2021} presents a more comprehensive overview of how ML has been applied within the field of material design in general. These developments are covered with great optimism, but the authors also highlight the common treatment of ML-models as black-box solvers for complex problems. Related to this, the review also observes that treating material design as image-to-image mappings, similar to the direct design applications for structural TO in the previous section, is widespread within this field. 

\paragraph{Multiscale TO}

{Multiscale TO (MSTO) is the approach of obtaining both the optimal structural topology on the macro-level as well as the local microstructure
material layout (\citealt{WuSigGro21}). In fact, this was the approach used in the original works on topology optimisation by \cite{BendsoeKikuchi1988} and  \cite{Bendsoe1989}. In \cite{BendsoeKikuchi1988}, effective properties of near-optimal rectangular hole microstructures were precomputed and interpolated, whereas they were computed analytically for optimal so-called rank-n microstructures in \cite{Bendsoe1989}. Recently, several works have used AI to learn the effective properties of various micro-architectures, effectively replacing the interpolations or analytical expressions from the original works (\citealt{KimLeeYoo2021, Whiteetal2018, Yilinetal2021, ZheKumKoch2021, WangBeckDa2021, Elingaard2021, ChandraSuresh2021, ChanDaWang2021, WangLiuDaChanChenZhu2021}). With such off-line computations, either analytical, interpolated or learned, very efficient MSTO algorithms can be constructed. The challenge of ensuring connectivity between local microstructures is taken care of by imposition of periodicity, which results in simple geometries but possibly deteriorated performance (\citealt{WuSigGro21}), by mapping techniques (\citealt{KimLeeYoo2021,ChanDaWang2021,WangLiuDaChanChenZhu2021}) or by special microstructure parameterisations that by construction satisfy connectivity (\citealt{ZheKumKoch2021,WangBeckDa2021,Whiteetal2018,Yilinetal2021}), but do not necessarily meet the theoretical bounds.}\\

\cite{KimLeeYoo2021} and \cite{ChanDaWang2021} utilised an ANN to model material properties for functionally graded composite structures. \cite{Whiteetal2018} utilised an ANN to model the elastic response of the microscale material within a gradient-based MSTO framework where the parameters describing the local microstructures were used as design variables. \cite{Yilinetal2021} similarly proposed a CNN for predicting the effective elasticity tensor and its gradients for voxel-based non-parametric microstuctures, and \cite{ZheKumKoch2021} for spinodiod {microstructures}. \cite{WangBeckDa2021} proposed a data-driven TO approach to multiscale cellular design for natural frequency optimisation including multiple choices for microstructure classes. By defining a finite set of different material choices, \cite{ChandraSuresh2021} integrated a multi-material blending scheme in the previous ANN reparameterisation approach for TO (\citealt{ChandrasekharSuresh2020}). \cite{Daetal2022} used an ML-inspired sampling procedure to construct a {database} of {microstructure} unit cells integrated in an approach to produce connected microstructures for indirect control of fracture resistance. \cite{WangLiuDaChanChenZhu2021} trained an ANN as a surrogate for modelling the geometry-property relation for parameterised microstructures to avoid homogenisation analysis during optimisation, while \cite{Elingaard2021} utilised a CNN as a surrogate for conventional de-homogenisation procedures. \\ 

The idea of using AI approaches to provide effective properties for multiscale approaches seems promising, especially for more complex non-linear problems where {CPU-heavy} path-dependent microstructure simulations would render a full multi-scale approach extremely expensive. Here, a costly off-line training at the microstructure level will be compensated in the form of much more efficient overall modelling and optimisation procedures.

\section{Assessments}\label{sec:2}
An overview of the works implementing AI-methods for use in TO was given in Section \ref{sec:1}. Within each of the presented application-categories several challenges and disadvantages were identified and discussed. This section will elaborate on some of these issues. First the importance and requirements for computational costs and generality of method applicability will be discussed followed by an assessment of the quality of solutions obtained and presented in the literature.

\subsection{Computational cost and applicability}\label{sec:2.1}
Disregarding the quality of obtained solutions, this section will focus on the overall merit of the different approaches in terms of the generalisation ability and associated computational costs. The motivation behind this focus is that the range of problems for which a model is applicable and the computational effort associated with generating training samples, running the learning algorithm and applying the proposed procedure to obtain a solution are factors strongly influencing the actual usefulness of the suggested framework. If a solution method is computationally expensive to prepare and tune and can only be applied to very specific problem cases, as seen for most of the direct design applications, it is not likely to provide any benefits regardless of the solution quality. Whether the structural results are promising should only be a determining factor in the evaluation of a method that offers a sufficient balance of speed-up and generality.\\

The \textbf{computational cost} of a method is not only related to the actual solution time, but time spent on collecting data and training might also have a significant impact as the computational cost of a conventional iterative TO solution procedure is typically what AI-technology is aimed at reducing. For instance, \cite{NakamuraSuzuki2020} sampled 333,000 TO optimised structures as training and validation data. This means that their resulting direct design model should be applied at least 333,001 times to similarly sized problems for any actual speed-up to be gained. Due to the need for target samples, examples of good quality solution structures to different problems, this holds true for any direct design model developed utilising a (semi-)supervised training approach.\\

If the range of problems the trained model is able to handle is extensive, an expensive training process becomes less concerning, but should still be a factor. Given a very large dataset, even a naive ML-model may perform well on a range of new problems. Consider a model which simply finds the example from the training data which is the most similar to the new problem, and returns this example's corresponding solution. In this case, the larger and more diverse the database known by the model, the better it will perform for most test cases. However, if a new problem has a significantly different optimal solution than any of the training samples, a new {data sample} similar to the new problem must be obtained by optimisation and added to the model's database before it is able to solve the problem. Thus, the model is not able to learn such that it can predict anything new. {Indications of this behaviour could be seen for the direct design model for fixed supports proposed by \cite{YanZhangXuetal2022}, where the predicted results for the test-cases resemble those optimised by SIMP in overall composition, but some of the predicted solutions to test instances were missing material where the load was applied. This could indicate that the test case corresponded to a small shift in {load locations} compared to one of the training instances, resulting in an infeasible solution, and implying that the model had not learned the significance of applied load positions.}\\

Considering the different application categories the threshold for achieving actual speed-up can be reduced in that the data needed is of a different, easier-to-obtain, nature or by increasing the degree of unsupervised learning when training the model. There are, however, still some non-direct design approaches that require computationally expensive training. \cite{Linetal2018} and \cite{SosnovikOseledets2017} utilised a direct design-like model to directly map from an intermediate structure obtained by SIMP to a converged structure. Thus the proposed approaches also require training data consisting of a large number of conventionally optimised structures, and with the added computational cost of performing {SIMP iterations} to reach the intermediate designs for new problems, not much is gained in terms of solution time speed-up.\\

\cite{KesAliTas2021} suggested an approximate way of computing the breakeven threshold for how many problems and AI-based solution framework must be applied before any actual gain in speed-up is obtained. This measure is obtained based on the computational cost in terms of FEAs needed for both obtaining the training data and executing the proposed procedure. It is not an exact representation of the computational costs as the use of optimisation algorithms in both the training of the network and the completion of the procedure may infer additional notable costs. Further, the computational cost of {FE-analysis} is computed as an approximative function of the number of elements in the FE-grid. The considered formula does, however, allow for comparing several different approaches and problem sizes in a relatively fair manner. \\

Let $\tau \in \mathbb{N}^{+}$ denote the desired breakeven threshold (i.e. the number of optimisations that must be performed before the considered method outperforms conventional TO methods) and  $C_{train}$ the computational cost of training. Given the computational cost of solving a problem over {an} FE-mesh of $N\in \mathbb{N}^{+}$ elements using both SIMP, $C_{SIMP}(N)$, and the proposed framework, $C_{AI}(N)$, the threshold $\tau$ can be computed as in equation (\ref{eq:break_thresh}).
\begin{equation}\label{eq:break_thresh}
    \tau = \dfrac{C_{train}}{C_{SIMP}(N)-C_{AI}(N)}
\end{equation}

For the direct design models reviewed in this paper, the breakeven threshold is commonly equal to the number of training samples used. This is because the mesh size $N$ is approximately (or exactly) the same for the training samples and the {problem formulations} for which the model is applied and, as FEA is not needed to execute the suggested AI-framework, one obtains $C_{AI}(N){=}0$. This {mesh dependency} is not represented in the given threshold computation, but is still a significant factor in terms of determining whether a proposed solution framework will be a reasonable alternative to conventional methods. If, for instance, a method that has a medium-level breakeven threshold, is restricted in terms of mesh size and dimensions and is focused on a limited set of boundary and load conditions, the perceived breakeven threshold might actually be much higher. This is because with limited applicability, there might not exist more than $\tau$ relevant problem cases that one could ever wish to solve.\\

Such considerations relate to the generalisation ability of a method. Judging the exact generalisation ability of a method can be challenging, but the aim should be to develop solution procedures that are as close to universal in terms of {problem definition} as well as {mesh dimensions} and resolutions as possible. Ideally the method should at the least be applicable to different sets of loads, boundary conditions length-scales etc. within the current problem setting. Further, it would be beneficial if the model is easily extendable to different objective functions and constraints as well.\\

Fewer limitations of {model applicability} imply greater generalisation ability. However, it is not necessarily expected that a useful application offers a universal solution approach to any problem. This is likely neither realistically achievable nor a quality found in conventional methods. The traditional TO methods are {inexact} in nature, relying on many different parameter settings, filters and search algorithms. One chosen combination of these is not likely to achieve universality and successfully solve any imaginable problem formulation. It is however possible to fit most problem formulations to the required ``inputs'' for conventional methods, and in many cases one can achieve satisfactory results with little parameter tuning. It is therefore fair to expect that e.g. a change of physics or constraints require retraining. However, for the purpose of pure compliance optimisation, one should assume that a single model setting is realisable. Generalisation ability can therefore relate to both the direct applicability of a method and its transferability. Transferability refers to how easily the model can adapt to new problems by changing parameters or network architecture, while still achieving good results.\\

To present an idea of how computational cost and generalisation ability compare for some of the key concepts presented in the literature, the breakeven threshold (\ref{eq:break_thresh}) from \cite{KesAliTas2021} is first adapted to approximate computational costs for a greater variety of applications.\\
For computing the breakeven threshold an estimation of computational cost associated with performing one FEA for a problem meshed using $N$ elements is defined by $C_{FE}(N)=N^2$. This is based on the computational order of $O(\mathit{bw}^2m)$ associated with solving a linear sparse system with a {$m \times m$} coefficient matrix and bandwidth $\mathit{bw}$. Assuming that for a 2D linear elasticity FE-problem discretised using 4-noded rectangular elements with equal number of elements in the $x$- and $y$-direction, $n_x=n_y$, the number of equations is given $m=2(n_x+1)(n_y+1)=2(n_x+1)^2$. Further, the best case node-numbering of such a mesh achieves a bandwidth of $\mathit{bw}=n_x+1$, and the number of overall elements $N$ is given $N=n_x^2$. Thus, the process of solving this system is of computational complexity $O(\mathit{bw}^2m)=O((n_x+1)^4)=O(N^2)$, which is used directly as the estimate for completing one FEA.\\

\begin{table*}[ht!]
    \centering
    \caption{\textcolor{black}{Components of method generalisation ability described in levels of achievements. Depending on model applicability for different BCs, {mesh dimensions} and loading conditions the levels are used as indicators for computing an overall generality score. The last category is related to how similar the training and test problems presented are. The last row indicates the weight prescribed to each category when computing the total score which is the weighted sum of the levels across the four categories, resulting in a generality scale ranging from 0 to 36.} {Note that computing the generality score for a given method does involve subjective assessment. Therefore, this score is not intended as a precise way to compare two specific methods. Rather, the aim is to provide a measure of how application categories compare and provide an illustration of overarching trends.}}
    \label{tab:Gen_calc}
    \begin{tabularx}{\linewidth}{c|LLLL}
    \toprule
        \makecell{\textbf{Category}/\\\textbf{Level}}& \textbf{Supports} & \textbf{Mesh} & \textbf{Loads} & \textbf{Test$\approx$Train} \\\midrule
         0 & fixed & small fixed & single-point restricted & sampled from the same limited pool\\\midrule
         1 & few (2-3 or very similar) & larger fixed & single-point many options & small difference (i.e. changing supports marginally resulting in small visual difference) \\\midrule
         2 & some (multiple options, but still similar problems) & limited variation possible (i.e. 2-3 choices for different aspect-ratios or resolutions) & multiple limited (i.e. $\leq$10 loads or placements restricted domain boundary) & some difference (i.e. definite difference aspect-ratios or loads)\\\midrule
         3 & many (a set of clearly different options) & many (some flexibility both in terms of shape and resolution) & multiple many (either no upper bound on no. loads or no restriction on placements) & medium difference (i.e. ensured difference loads and BCs) \\\midrule
         4 & any (method appears applicable to any set of reasonable support options) & any (complete mesh-independence) & any (no restrictions apply) & significant difference (BCs, loads, mesh-shape and resolution different) \\ \midrule
        \textbf{Weight} &\multicolumn{1}{c}{3} &\multicolumn{1}{c}{2} &\multicolumn{1}{c}{3}&\multicolumn{1}{c}{1}\\\bottomrule
    \end{tabularx}
\end{table*}

Let $\overline{N}$ denote the set of considered mesh resolutions, $s^{method}_N$ the number of samples at mesh size $N\in \overline{N}$ needed by the method and $t^{method}_N$ the average number of FEAs the method uses to obtain a sample at mesh size $N$. Based on these definitions the computational cost of a method can be computed by (\ref{eq:cost_method}).
\begin{equation}\label{eq:cost_method}
    C_{method}(\overline{N})=\sum_{N\in \overline{N}}s^{method}_N\cdot C_{FE}(N)\cdot t^{method}_N
\end{equation}
The computational cost of obtaining the training dataset can thus be computed given the different {mesh sizes} considered, the needed number of samples for each {mesh size} and the cost of obtaining a sample at each {mesh size}. $C_{train}$ is as such the sum of the computational cost associated with obtaining each training sample, while $C_{SIMP}(N)$ is the cost of the average number of FEAs needed by SIMP to solve a given test problem of size $N$. $C_{AI}(N)$ depends on the specific framework considered. For direct design models this cost is zero, for post-processing upscaling methods this cost usually corresponds to $C_{SIMP}(\alpha N)$ where $\alpha<1$, while within the acceleration category this computation varies considerably between the individual applications. As this measure is still approximative it is not the exact {threshold value} that should be of interest, but the order of magnitude which is believed to be appropriate even with different estimations for $C_{FE}(N)$.\\

For quantifying the \textbf{generalisation} ability of the applications no corresponding measure is available. Therefore a set of criteria are developed to manually judge the perceived generalisation ability of an application based on method description and presented results. These criteria are listed in Table \ref{tab:Gen_calc}, and relate more to the range of possible problem applications than the actual performance in terms of solution qualities. Poor results, in terms of structural disconnections or difficulty adhering to the constraints, do in practice have an influence on the generalisation ability, as the models should be able to feasibly solve the intended test problems. However, to limit the level of subjectivity in the generalisation score, this is not accounted for.\\

The generalisation ability for a method is calculated as a weighted sum of the levels across the four categories in Table \ref{tab:Gen_calc}, and is as such subjectively judged on a scale from 0 to 36. From the given weights, the main determining factors are {mesh dependency} and the variety of viable problem definitions the application can handle (category 1-3). As such, a solution framework is appointed a low generalisation ability if it is developed for a fixed FE-mesh with training and test problem instances limited by fixed boundary conditions, while a high score is allotted if there is no {mesh dependency} and the application is believed to work for any reasonable boundary conditions. The fourth category relates to the similarity between the training and test data-sets, as to penalise instances where no direct effort is put towards making these clearly distinguishable. Generally, this is a very important factor for any ML model, as overlaps between test and training data are likely to lead to inflated performance measures, overestimating the capabilities of the model. This factor is still given a lower weight in this specific assessment, related to the choice of differentiating more between generality and solution quality. {The overall purpose of this quantification of generalisation ability is not to serve as an absolute and true instrument for evaluating AI-based solution methods in TO, but rather as a measure for illustrating the current state of the field and how the application categories compare.}\\

%\begin{figure}[ht]
%    \centering
%    \includegraphics[width=\linewidth]{Fig5_generalisation.pdf}
%    \caption{Comparing, for different methods in the reviewed literature, the perceived generalisation ability along the x-axis to when speed-up is achieved, by the breakeven threshold $\tau$, along the y-axis. The datapoints are coloured based on the application category to which they belong. }
%    \label{fig:sec21_gen_vs_cost}
%\end{figure}

\begin{figure*}[ht]
\centering
\includegraphics[width=0.9\linewidth]{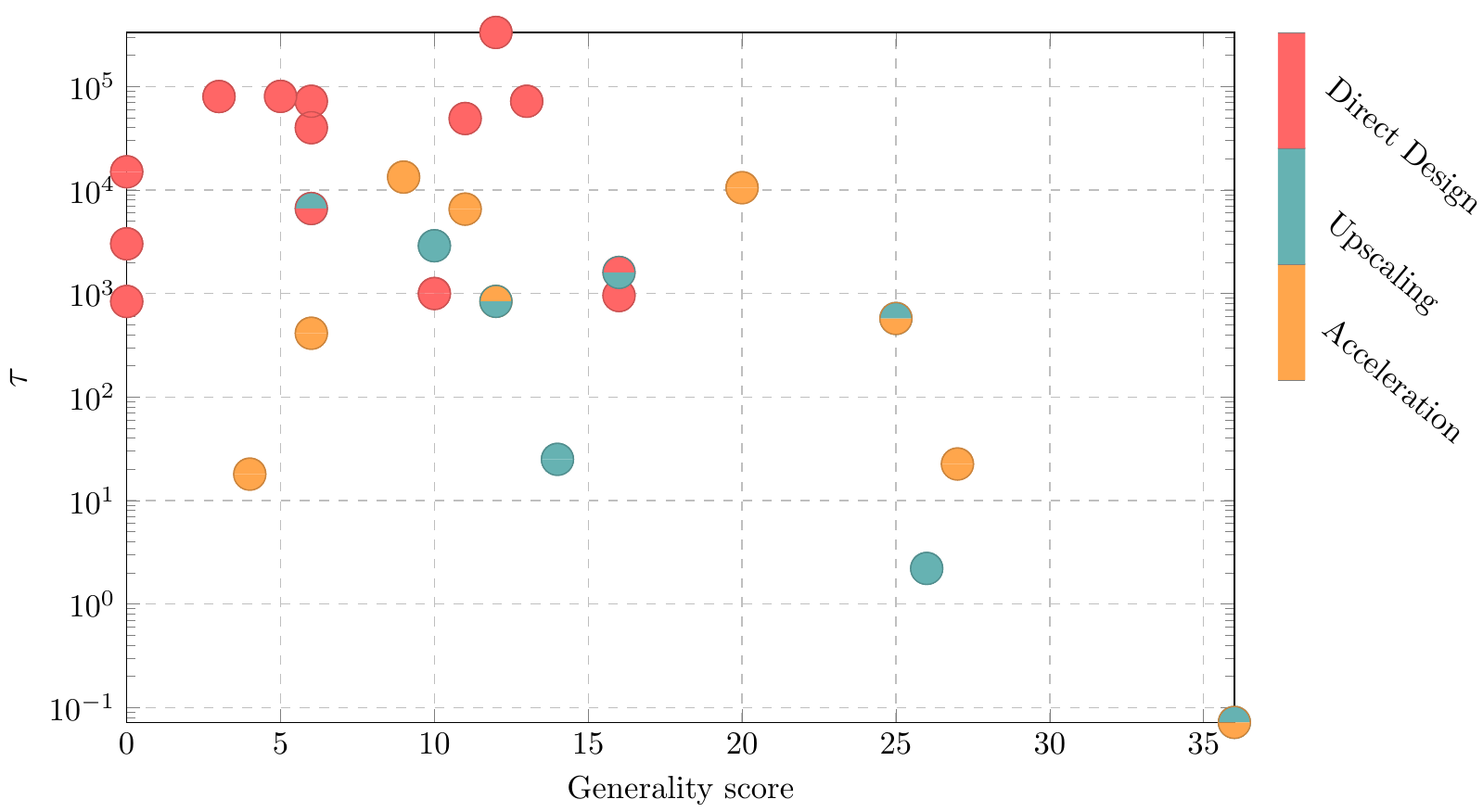}
\caption{Comparing, for different methods in the reviewed literature, the perceived generalisation ability along the $x$-axis to when speed-up is achieved, by the breakeven threshold $\tau$, along the $y$-axis. The datapoints are coloured based on the application category to which they belong. }
    \label{fig:sec21_gen_vs_cost}
\end{figure*}

Fig. \ref{fig:sec21_gen_vs_cost} illustrates the breakeven threshold compared to the perceived generalisation ability for a selection of methods in the literature reviewed. It is noted that only literature providing sufficient information to estimate the computational cost is considered. Further, only methods that are directly comparable to conventional iterative solution methods like SIMP are included. Therefore the presented applications belong to the direct design, acceleration or upscaling categories, which is illustrated by the different marker colours. The dual-coloured {scatter points} indicate that the associated solution framework is a hybrid between the two corresponding categories. For instance, \cite{LiBetal2019} used a direct design model to directly predict an optimised low-resolution structure before applying upscaling to achieve the final desired structural layout. \cite{ChiZhangetal2021} and \cite{KalliorasLagaros2020} both utilised upscaling as an integrated part of the acceleration-based iterative procedures proposed.\\

As both presented measures, breakeven threshold and generalisation ability, are approximative, the datapoints are in general not explicitly linked to the corresponding works in the presented figure. It is however evident that solution methods belonging to the direct design category perform worse than the other alternatives as seen by their high breakeven thresholds and low generalisation abilities. The best scoring direct design model (\citealt{ZhengHeLiu2021}) is given a higher generality score than the other similar approaches due to the incorporation of zero-padded buffers in the network input format to allow for different {mesh dimensions and resolutions}. However, due to the only guaranteed difference between training and test data being the {mesh size} and the limited sample space for problem characteristics, it is still seen as having a low generalisation ability. The lower breakeven threshold is obtained by considering overall finer {mesh discretisation} for the test problems than for the training problems. Even though 7,500 optimised structures are used for training, the breakeven threshold is approximated to be lower than 1,000.\\

For the acceleration and upscaling categories the performances are more varied across the different approaches, where worst case is similar to the poorest of the direct design models while best case approaches a negligible breakeven threshold and high generalisation score. Two acceleration-based methods are given the lowest generality score, both due to the restricted problem spaces considered. The first (\citealt{Linetal2018}) restricts the problem to a small, fixed mesh, 2D thermal conduction problem where only the volume fraction, heat source and sink are varied. One of the two highest breakeven thresholds for the considered acceleration-methods is also reported for this model, linked to the use of SIMP paired with a direct design-like model mapping an intermediate solution to a final design. The second (\citealt{QianYe2020}) has a significantly lower threshold because ANNs are used for objective and sensitivity predictions within a conventional SIMP framework, such that fully optimised designs are not needed for training. In terms of generality, the model is, however, trained and tested considering a set of fixed boundary conditions with different single-point loads, resulting in a poor performance.\\

The overall best scoring work is that of \cite{ChiZhangetal2021} which has a breakeven threshold close to zero and is given a generality score of 36. As the method is based on a concept of online learning integrated in a two-scale optimisation procedure no cost is associated with pre-training of the model and the overall computational cost to achieve an optimised structure is lower than that of conventional methods, and thus a speed-up is achieved regardless of how many times the model is used. This approach is also what makes the degree of generality so high, as the network input and output format can be adapted to each individual problem. \\

There are some promising works that exhibit great generalisation abilities but are not included in the evaluation in Fig. \ref{fig:sec21_gen_vs_cost} due to deficiencies in reported computational costs. This holds true for most of the reduction-category frameworks. The relevant works apply an ANN for reparameterising the geometric representation and optimise a given topology through online training of the network. This means that the network model in itself is independent of the boundary conditions, but adapts to the given problem during optimisation similarly to a conventional TO process. As such, there is little to no computational cost associated with offline {data collection} and network training, contributing to a lower breakeven threshold. Further, the generality of these frameworks is expected to be high in terms of solving different problem formulations. \cite{ZhangZhao2021} exemplified this by applying their proposed method to minimum compliance problems (both with and without {stress constraints}), compliant mechanism design and a 2D heat conduction problem. An extended framework for nonlinear elasticity was also proposed and tested. The complications with assessing such reduction methods arise from the fact that FEA is still needed for the full mesh in each iteration, as it is required for evaluating the {loss function} of the ANN. As FEA is the most expensive operation in conventional topology optimisation (and used here for approximately measuring speed-up), the gain in computational efficiency of such methods rely on a reduction in the number of iterations needed to reach convergence. Such behaviour is observed for some selected comparisons to SIMP reported in the literature, but no extensive analysis allows for conclusive judgement.

\subsection{Solution quality}\label{sec:2.2}

After having assessed the merit of the different solution frameworks in terms of generalisation ability and computational cost as in Section \ref{sec:2.1} evaluation of solution quality in terms of structural performance should be included as a third dimension for judging overall performance. Overall it is deemed that any model with higher breakeven threshold and lower generalisation ability should be disregarded regardless of solution quality. However, as most of the works represented by data-points located in the top left of Fig. \ref{fig:sec21_gen_vs_cost} describe direct design models, it is worth noting the problematic solution qualities discussed in Section \ref{sec:1.1} and the common occurrence of disconnections as in Fig. \ref{fig:test1}. \\

For the remaining works, for which the combination of generalisation ability and breakeven threshold is considered reasonable, the solution quality and speed-up are the integral measures of quality. Speed-up here refers to the actual solution time, and thus, provided a trained model and solution framework, the reduction in computational cost of solving a problem compared to conventional state-of-the-art methods. Therefore, it would be desirable to evaluate methods in a similar manner as in Fig. \ref{eq:break_thresh}, but now for mechanical performance and solution time.\\

It is found that a majority of the papers do not present performance metrics allowing for fair quantifiable comparison of results. There are several different reason for the presented results not being suitable for assessment. Firstly, there is a tendency to only present the mean value of the selected performance metrics for the test instances, or to only present the results for a few illustrative problems from the test set. Such formats for reporting results may not be representative {of} the full distribution of performance across the entire set, potentially hiding outliers for which very poor performance is observed. Secondly, the overlap between the training and test sets may be large, by i.e. randomly sampling from the same restricted pool of problem combinations, which is likely to overestimate the performance and not be representative if different problem characteristics are considered. Thirdly, the choice of performance measures are often related to pixel-wise density errors only, neglecting to evaluate the actual structural properties of the obtained structure. Lastly, the comparison to conventionally obtained benchmarks is not done in a fair manner. Many works compare the obtained results, in terms of relevant measures such as structural compliance, to {grey scale} structures obtained by SIMP using a large filter radius. It was shown in Fig. \ref{fig:test1} how compliance is reduced significantly when the structure is thresholded to a solid-void design. Further, a large filter radius means that the structure will contain fewer fine features, which may influence the structural performance of the resulting design negatively\footnote{\cite{SigmundMaute2013} included a few lines of Matlab code to perform a volume-preserving threshold.}. Also, as several of the proposed frameworks do not contain mechanisms for directly enforcing the volume constraint, this is also an important property, as more material results in lower compliance. Inspired by these challenges, and to allow for future fair comparisons of quality, recommendations for how to test the model will be given in Section \ref{sec:3.2}.\\

Many applications within the upscaling and acceleration categories exhibit similar problems in terms of the solution quality assessment, based on the reported experimental results. Firstly, very few coarse grid cases are considered for testing, and in some cases they are very similar to training. Secondly, the reported results tend to focus on average {elementwise-density errors} which means that quantitative assessments of actual structural performances and expected worst case behaviour are prohibited. Generally, illustrations of the obtained structures, for selected test-instances, using both conventional and AI-based methods do, however, show a high degree of {grey scale} and tendencies for disconnections. The frameworks including features of upscaling tend to yield blurry boundaries resulting in structures without the fine features of those obtained by conventional methods, due to length-scale restrictions provided by the coarse mesh. \cite{Xueetal2021} showed the only noticeable occurrence of the AI-based method obtaining a structure with finer features than those obtained by SIMP. These structures also exhibited lower compliances than the corresponding benchmarks, but as the benchmarks were obtained by enforcing a large filter radius, it is unclear whether the results are better than what can be obtained by SIMP overall.\\

\begin{figure}[ht]
    \centering
    \begin{minipage}{1.0\linewidth}
    \includegraphics[width=\linewidth]{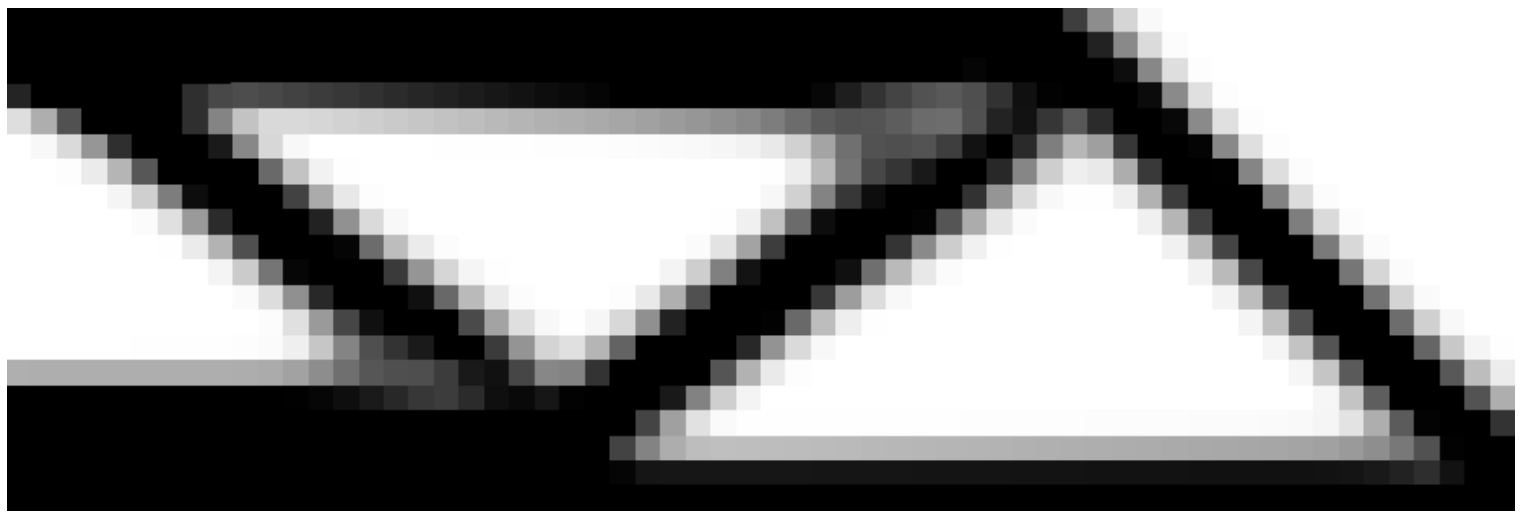}
    \subcaption{{Optimised MBB-beam on a 60x20 FE-mesh w. volume fraction 0.5 and filter radius 2.0 (Compliance 209.1529).}} 
    \includegraphics[width=\linewidth]{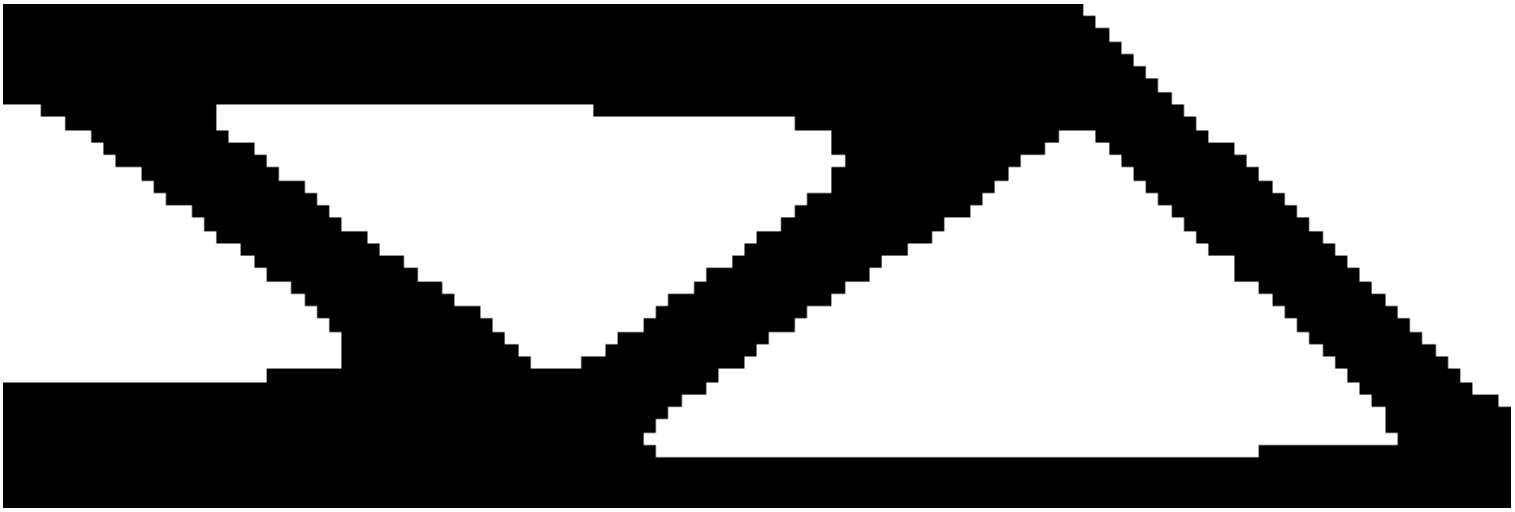}
    \subcaption{{Upscaled and projected MBB-beam to a 120x40 FE-mesh w. compliance 191.1536 (compared to 191.2364 in \cite{ZhangZhao2021}).}} 
    \end{minipage}
    \caption{A solution to the MBB test case from \cite{ZhangZhao2021} obtained by optimising on a coarser grid (a), resizing the image using Matlab built-in function with bicubic interpolation followed by volume perserving thresholding (\citealt{SigmundMaute2013}) to obtain higher resolution solid-void structure (b).} \label{fig:test3}
\end{figure}

The reduction category applications are typically tested on a wider range of problems and {mesh discretisation}s, but the results presented tend to lack transparency as for the above-mentioned methods. A clear benefit is that the obtained structures appear to have a smoother and sharper structural representation than the presented solutions within the other highlighted categories. As regular FEA is used to extract the structural geometry for analysis, this feature is, however, not utilised in a beneficial manner. What is also prominent is the loss of finer features, compared to conventionally obtained solutions to the identical problems. This loss in detail could, in some cases, be a consequence of the reduced geometric representation not being able to describe the more complex features. A reason for fewer iterations needed in these reduction-based procedures could therefore be related to the size of the feasible solution space. Optimised structures illustrated in \cite{DengTo2021} do indicate that capturing the finer features is achievable with a larger number of hidden layers. \cite{ZhangZhao2021} argued that the thicker structure with fewer holes is an advantage of their proposed method as these properties make the structure easier to manufacture. Here it is relevant to note that one easily can reduce the fine features resulting from SIMP by increasing the filter radius or lowering the {mesh resolution}, and that several strategies for achieving length-scale control and manufacturability have been reviewed in \cite{LazarovSigmund2016}. {In fact, for the MBB-beam example presented in \cite{ZhangZhao2021} a similar structure with a lower compliance can be achieved by considering a coarser mesh when applying SIMP for optimisation (Fig. \ref{fig:test3}(a)), and then upscaling the optimised structure using the \verb|imresize| image-scaling procedure in Matlab with bicubic interpolation, before applying a volume preserving threshold to obtain a 0-1 design (Fig. \ref{fig:test3}(b)). The corresponding Matlab-code for performing this sequence of operations is found in Appendix \ref{app:sec1}. Doing so decreases both the computational cost of FEA in each iteration, as well as the number of iterations needed (Fig. \ref{fig:test3}).}

\section{AI limitations}\label{sec:666}
State-of-the art ANNs have achieved a level of pattern-recognition abilities exceeding human abilities. This development has been made possible by the availability of massive sets of labelled data and increased computing power. High-level ANNs contain tens or hundreds of hidden layers and are trained utilising high-performance GPUs and hundred of thousands to millions of data samples. For instance, AlexNet, the first ANN to adopt an architecture consisting of consecutive convolutional layers and a leading model for object-detection, was trained on 1.2 million labelled {data samples} and has obtained the ability to recognise 1,000 different objects (\citealt{Krizhevskyetal2017}). An ANN learns directly from the provided data with limited human influence on what it learns. These models are powerful tools because they allow for a complexity enabling pattern recognition that is not explicitly defined by human understanding. Successful application of these models do, however, often rely on carefully curated data, and the measured accuracy of an ANN is typically measured using standard benchmark datasets. The measured accuracy depends greatly on the nature of the chosen test data, where greater similarities to the training data implies a higher measured accuracy (\citealt{Goodfellowetal2018}). \\

The current leading deep learning technology is fundamentally brittle, as it breaks in unpredictable ways when exposed to unfamiliar domains (\citealt{Heaven2019}). A small perturbation or added noise to input samples may lead to incorrect predictions with high confidence and the same input is able to break a wide variety of {model architectures} trained on different datasets (\citealt{Kurakinetal2016}). Such failures are strongly related to ANNs not actually understanding the world or comprising any knowledge of salient features, which are core factors in human recognition. ML models are traditionally developed under the assumption that the environment it is operating within is benign both during training and validation and that the sample-space distribution is the same as for the training data in any future tests. There exists measures for increasing the robustness of ANNs in form of data augmentation (\citealt{Goodfellow2016}) or adversarial examples (\citealt{Kurakinetal2016}), but increasing robustness against one type of error could weaken the model against others. It is commonly believed that augmentations to the current model-framework providing additional reasoning abilities could aid in overcoming this brittleness (\citealt{Heaven2019}). \\

When utilising these frameworks to solve different tasks it is therefore important to be realistic about what the possible capabilities of any ML-model are. Many of the articles included in this review appear to overestimate the abilities of current ML-technology, ignoring that these models simply are complex versions of regression and classification and treating them as some black-box ``magic'' solver able to handle any complex task.\\

\begin{figure}[ht]
 \begin{minipage}{0.65\linewidth}
     \captionsetup{width=1.0\linewidth,justification=raggedright,singlelinecheck=false}
         {\includegraphics[width=0.95\linewidth]{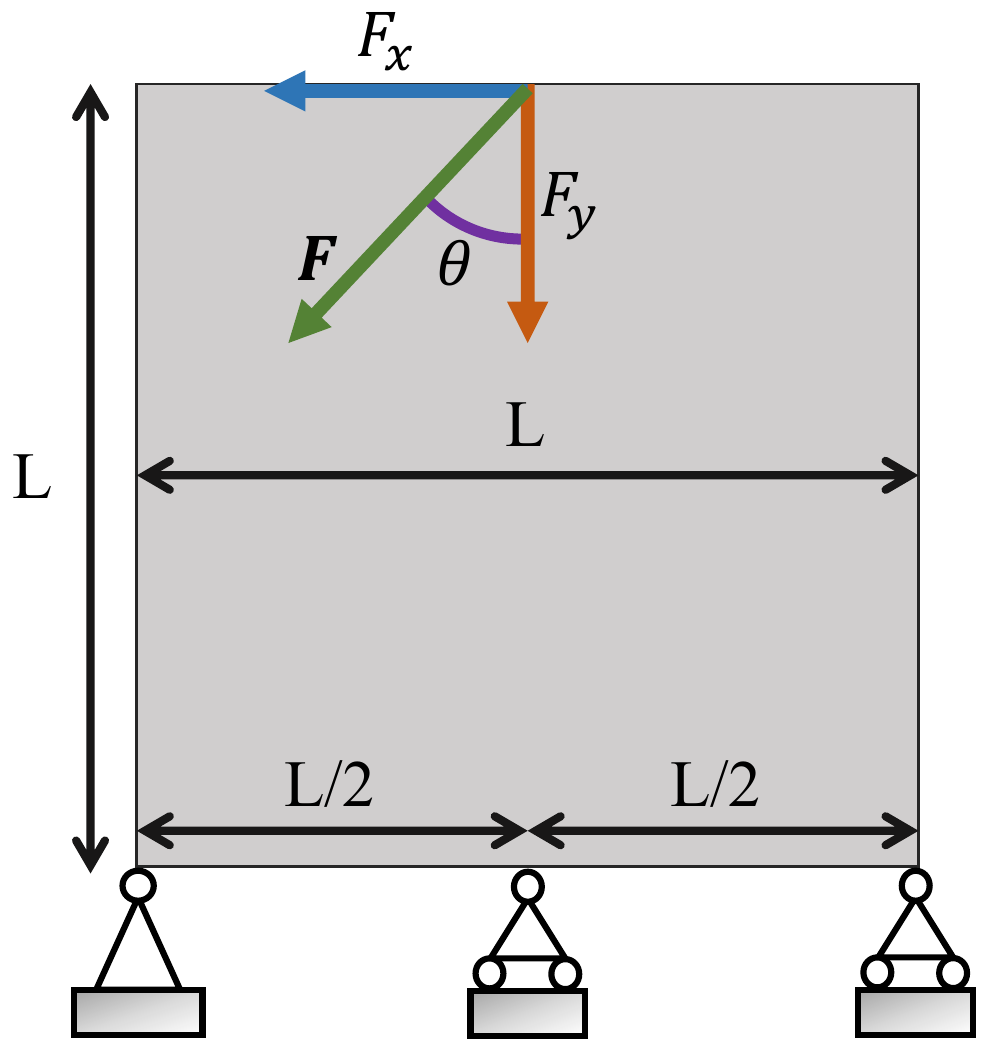}}
         \subcaption{{Load magnitude $\lvert \mathbf{{F}}\rvert{=}1$, {volume fraction} of 0.3, 64x64 elements, filter radius 2.0}}
         \label{fig:test2_BC}
 \end{minipage}
 \begin{minipage}{0.33\linewidth}
          {%
        \setlength{\fboxsep}{0pt}%
        \setlength{\fboxrule}{0.5pt}%
        \fbox{{\includegraphics[width=1.0\linewidth]{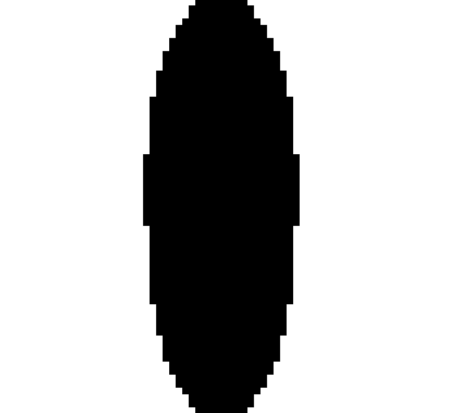}}}%
        }%
         \cprotect\subcaption{$\theta{=}0$}
         {%
        \setlength{\fboxsep}{0pt}%
        \setlength{\fboxrule}{0.5pt}%
        \fbox{{\includegraphics[width=1.0\linewidth]{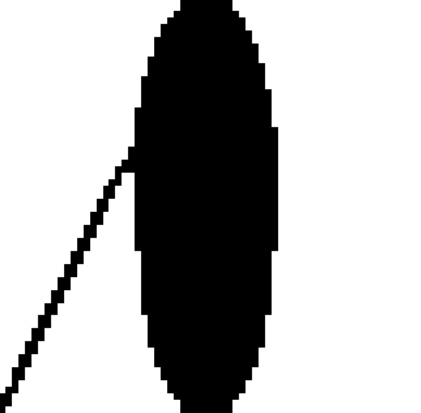}}}%
        }%
         \cprotect\subcaption{$\theta{=}0.01$}
 \end{minipage}
      \cprotect\caption{Test-case illustrating the potentially significant effect small changes in boundary conditions can have on the optimal material layout. The boundary conditions (a) with corresponding optimised 0-1 designs for (b): $F_x{=}0,\;F_y{=}1$ and (c): $F_x {=} 0.0099998,\;F_y {=} 0.99995$.}
      \label{fig:test2}
 \end{figure}

The direct design approaches are often based on generative methods (e.g. GANs or VAEs). The assumption of such methods is that the output should belong to a specific distribution, and generative methods aim at generating outputs (for instance a 2D or 3D image) that belong to such a distribution. In practice, generative methods, and most CNN-type networks, often learn a mapping from a latent space of much smaller dimension to the output image. It might seem reasonable to assume that one could simply interpret (or map) the boundary conditions for a TO problem as (or to) a point in the latent space which is subsequently mapped directly to a mechanical structure using a mapping that is learnt from examples. Given sufficient examples, one might expect this to work. Unfortunately, there are good reasons to believe that such strategies will always fail. The main reason is that relatively small changes to the boundary conditions can lead to a very different solution being optimal. Thus, any learning based approach would face the challenge that a small perturbation of the boundary conditions could lead to a big change in the optimal structure. Unless the types of {problems that} can be solved with a hypothetical direct design approach based on a generative method, are strictly limited, it is clear that an unbounded number of examples could be necessary to learn all the discontinuities in the mapping from latent space to mechanical structure. An example of this is illustrated in Fig. \ref{fig:test2}, where the problem described in Fig. \ref{fig:test2}(a) is optimised for two single-point load cases. The first case describes a vertically applied load, and the obtained structure is given in Fig. \ref{fig:test2}(b). For the second case the applied load is applied at a slight angle and the optimised structure is now given in Fig. \ref{fig:test2}(c).
 
 The difference between the optimised structure of the first and second considered load case is that the first is simply one vertical bar to counteract the applied load, while the second has an additional thin bar to also supply support in the horizontal direction. Table \ref{tab:test2_compliance} compares the compliances of these two optimised structures when they are subjected to each of the two load-cases. It becomes evident that the small load-angle perturbation results in a collapse of the first structure (Fig. \ref{fig:test2}(b)) while the second structure (Fig. \ref{fig:test2}(c)) performs similarly in both cases.
 
 \begin{table}[ht]
 \caption{The compliance of the optimised structures in Fig. \ref{fig:test2} (b) and (c) with respect to different $\theta$-values changing the loading conditions in (a).}
     \label{tab:test2_compliance}
     \centering
     \begin{tabular}{c|cc}
     \toprule
          Design& (b) & (c) \\\midrule
           $\theta{=}0$ &  7.255 & 7.372 \\
           $\theta=0.01$ & 1,649,351,760 & 7.5591\\ \bottomrule
     \end{tabular}
 \end{table}

Further arguments could be made about the individual element-density time-series acceleration approach proposed by \cite{Kallioras2020} also suffering from the lack of understanding of physical properties. For many problems, it is plausible that the iterative history of element densities will map to a reasonable structural composition, but this cannot be guaranteed for all problems. As the model does not know the relations between neighbouring elements there is no guarantee that even a smooth structure will emerge when combining the individual predicted trajectories. This is not to discourage that simple mappings between intermediate solutions in the optimisation process may aid in faster convergence, especially as {SIMP iterations} are run on the mapped structure to ensure physical consistency in the final design. One should however be cautious of the fact that this does not guarantee increased performance, and especially not for problems with moving members during the later iterations.\\

The approach of reducing the number of exact high-resolution FE-solutions in a multi-scale framework proposed by \cite{ChiZhangetal2021} might be a more rewarding application of approximate mappings. During several iterations of the TO process, approximate sensitivities may be sufficient to make progress in the overall optimisation, as also investigated for conventional TO (\citealt{AmirSigmund2011}). Therefore, understanding the problem may not be necessary to perform this task, and errors in the predictions are less likely to have major effect on the overall results. A fast increase in prediction errors as a function of the number of iterations since the last exact sample may indicate that model-improvements could be fruitful. \\

The upscaling category on an overall level likely suffers from similar lack of physical understanding as the direct design models. It is observed that the applied models mainly perform boundary smoothing, which is understandable if the mapping between resolutions is simply seen as an image-resizing task by the ANN. The work of \cite{Elingaard2021} is an outlier within this category, as here the {ANN} is utilised to perform de-homogenisation. As physical understanding is not necessary to translate the input vector-field to an intermediate density field the abilities of ANN to perform pattern recognition may be very useful in such an application. However, it is found that the post-processing procedure utilised could benefit from some {alterations} to replace the u-shaped branching with v-shaped ones, which are known to provide better structural performance.

\section{Recommendations}\label{sec:3}
The motivation behind implementing an AI-model to be used in TO should rest on some belief that it will improve the overall solution framework, compared to conventional procedures. As the limiting factors for large-scale TO are related to memory consumption as a result of the design representation and computational time associated with the calculations required in the applied iterative solution procedures, the aim should be to reduce the cost of at least one of these challenges without compromising performance with respect to the other. Scientific contributions are only significant if they increase the scope of viable applications. This could be solving any problem faster than conventional state-of-the-art approaches or extending the range of problems that can be solved, both in terms of size and complexity. \\
These statements are not to diminish the value of proofs-of-concept in a new application area of AI, but there should be convincing arguments and proofs supporting the capabilities of the presented ideas. Such considerations should focus both on how the method compares to current state-of-the-art and the range of problems that can be handled appropriately. New solution procedures should be accompanied by convincing descriptions of why they contribute positively to the scientific progress within the field. If insufficient results are obtained, a thorough reflection of why the motivating hypothesis or method-construction was wrong would be of great benefit for the field moving forward. \\

AI-technology has become an immensely popular research topic due to the prospective abilities of methods within the field. It is however important to recognise that the current capabilities of even state-of-the-art applications are limited and susceptible to errors. The {application area} of such models should therefore be robust with regards to possible {model errors} not having significant negative effects on the system as a whole. Further, the tasks at which existing technologies excel are limited and mostly related to pattern recognition. Methods such as neural networks are good at {data reconstruction} and feature extraction, but are insufficient when it comes to higher level understanding and data interpretation. This again is not to discourage research focused on applying AI-methods for TO. The combined choices of {model types}, model-configurations and applications are endless, and successful choices may provide meaningful contribution to scientific progress. To approach such useful choices one should however ensure understanding of the different {model choices} and practice critical evaluation of both the potential and longevity of the modelling framework itself, as well as the observed performance through testing.\\

Based on the ideas and results observed in the current literature, this section will therefore provide what is believed to be basic but useful guidelines for researchers interested in pursuing different application of AI in TO. To this purpose recommendations for what to consider when developing a model as well as how to properly assess its resulting performance are listed in the following.

\subsection{When designing the model}\label{sec:3.1}
{When choosing a model type} and its configurations the aim should be to determine which settings are likely to perform best for the intended application. With AI-methods this can be difficult as {model design} is not an exact science, but efforts should still be made to research different options and how they have successfully been applied in other fields. Doing so, it is also paramount to account for all aspects having an influence on the problem at hand, and remembering that relational properties should be provided to the model as one cannot expect the model to magically realise the environment it is applied within. Some concrete comments on the choices made in the current literature will highlight how such challenges can manifest, but as this very much is a new and open research field a complete guide cannot be provided.\\

After having identified the desired task, efforts should be focused on determining appropriate inputs, outputs and {loss function}. The analysis of breakeven threshold and generalisation ability in Section \ref{sec:2.1} illustrates the effect of data formatting on the computational performance of the model. It is found that in current works the training data generation requires a high computational cost which is often incorrectly deemed negligible. It is fair to assume that the more versatility the obtained model can provide, the more training time one can justify. However, most of the reviewed works having greater generalisation abilities typically present methods that have lower training costs. The reason for this is likely related to the fact that cheaper training data often is more removed from any specific problem instance, such that patterns the model learns from are relevant for a wider variety of cases. \\
The chosen model output is therefore a deciding factor in the {model design} phase. These outputs should be realistically attainable by the chosen model and preferably in a format that allows for a wide range of problems and {mesh dimensions}. Further, it is beneficial if benchmark targets can be computed somewhat efficiently as to evaluate the accuracy of the output. This means that neural networks aimed at directly predicting an entire {output structure} like the described direct design models might be unrealistic and very limited in terms of generality. Another limiting effect of using optimised structures to train the model is that it is likely to be influenced by the nature of the conventional solver used and the select set of parameters, where for instance using a larger filter radius to obtain the training samples may result in the model never considering finer features as optimal.\\

The formatting of the input and {loss function} {is also an important aspect}, as the information contained should, ideally, be sufficiently able to describe the unique properties of a considered problem instance, such that there is a believable relation between the input and the output. The input should allow for different problem dimensions while the {input dimensions} are kept limited to reduce the size of the network. Therefore it is recommended that greater efforts are made towards defining an appropriate {loss function} to capture more complex system relations.\\

These aspects means that one should seek more complex {loss functions} relating more to physical properties that require FEA in the evaluation of results and less to image-based prediction error. The norm should be to move away from designing mesh and boundary condition specific models and towards more adaptable frameworks than what is seen in most of the current literature. Further, it is recommended to consider automated post-processing as a valid option to refine the final structures to i.e. limit the occurrence of disconnections. In many cases this can be done in an inexpensive manner, such that the computational time is not significantly affected.\\
Overall, the most important reflection points are related to how many different problem types or settings the method can handle without retraining, while supplying good quality results.

\subsection{When testing the model}\label{sec:3.2}
When testing the model the aim should be to accurately determine whether the performance for the intended task is sufficient. This relates both to the statistics presented for model evaluation and the benchmarks used for comparison.\\
Fig. \ref{fig:test1} in Section \ref{sec:1.2} proved how the pixel-based density error is not an appropriate error measure when evaluating structural performance, as a small pixel-based error may imply good performance even for disconnected structures. Image-related {error measures} in isolation say more about the exactness of the reconstruction than whether the approach was successful in solving the TO problem. The physical performance of the obtained solution in terms of the desired properties defined by the optimisation objective and constraints should therefore be the main focus. When testing the model on several {problem cases} it is also crucial to include the entire distribution of the obtained results. This is to ensure that worst case behaviour is identified, because in structural design, even one insufficient solution can have detrimental consequences.\\

In terms of benchmarking it is both important to ensure the desired test cases are notably different from all cases considered during training and that the comparison to conventional methods is done under fair circumstances. Firstly, this means that it is not sufficient to test on random problems that are sampled from the same limited pool as the training data. Secondly, it should be assured that the two compared structures are defined on the same length-scale and thresholded to a black-and-white design, satisfying the same {volume fraction}. Some of these latter requirements might differ depending on the considered optimisation problem, but as most works are concerned with compliance-minimisation subjected to a volume constraint, it is convenient to base the recommendation on this type of problem.\\

If comparison to other similar AI-based models is conducted, all aspects should be considered and one should aim at ensuring the same testing conditions. As such, similar problems should be solved, the same performance measures should be used and training time, including {data generation}, should be included in the assessment. If a larger number of {data samples} or longer training time is allowed, increased prediction performance may be observed without the {model architecture} actually being better.

%\color{red}
\subsection{Benchmark cases}
The previous section exemplified how results should be presented and compared to conventional approaches for standard minimum compliance problems, and has encouraged proving performance for a wider range of problems. What has not been covered, however, is how appropriate problems for benchmarking should be chosen.\\

Constructing a set of specific benchmark problems that should be included to prove the validity of a conventional TO framework is difficult in itself, but for NN-based frameworks this becomes even more challenging as it also depends on how the NN is applied and what data is used for training. Firstly, if a model is trained on these specific benchmarks, then good performance for these problems is not an indication of the model's general capabilities. Secondly, the different ways in which ANNs can be applied within TO further individualises what appropriate benchmarking is. Therefore, instead of presenting a finite set of problems to include, this section will cover some important considerations when choosing the test-cases for a specific ANN-framework and how to evolve from a proof-of-concept to certified state-of-the-art legitimacy of the proposed approach.

\paragraph{The basic}
The first level of difficulty for benchmark problems can be directly related to the definition of generalisation ability in Table \ref{tab:Gen_calc}, where the aim should be to justify the highest score possible. An approach for achieving a high score could be to select a small set of domain shapes different from the training problems and for each of these shapes consider separate combinations of mesh-resolutions, supports and loads.\\

%What one should further consider within this basic framework is whether the length-scale is fixed, arbitrarily varying or can be controlled....???\\
Based on how the ANN is integrated within the optimisation framework, i.e. what application category the presented model belongs to, there may be additional considerations that become important when benchmarking. If the ANN is used to perform a sub-task in the optimisation process it is of interest to assess performance for this specific task, in addition to how the overall optimisation procedure is improved. This means that for ANNs trained to perform FEA the accuracy of the obtained sensitivities should also be assessed. Approximate sensitivities could lead to good quality optimised results, but the accuracy may still have effects on the reliability of the approach for future cases. Further, for convergence type frameworks mapping between intermediate designs to skip parts of the optimisation process, the output from the model, before continued iteration, is also of interest. This last point is of particular interest because if the achieved speed-up is the result of pixel-rounding or moving average type changes, one can avoid the expensive training and obtain more general frameworks performing equally well. \\

The issue of coarse-mesh restrictions on the minimum length-scale of high resolution structures obtained by two-scale applications has been a prominent topic throughout the review. Assessments of this effect should thus be included for these methods. One potential approach could be to illustrate the underlying coarse mesh of the fine scale result. Restrictions on the complexity of the obtained structures were also evident for re-parameterisation approaches due to the reduced design-representation and solution space. The question for such approaches could therefore be whether one can control the length-scale or impose local volume constraints to force fine features in the optimised structure (\citealt{WuAageetal2018}). \\

Lastly, there are some works claiming that the proposed NN-frameworks are justified only for problems where conventional approaches are less efficient or effective. In such cases it is crucial that the provided benchmark cases belong to the problem-categories for which the framework is intended to be beneficial.\\

The generalisation ability measure includes no requirements about changes in problem definition (material parameters, objective function, constraints) or the scale of the different mesh-resolutions. For developing new methods for TO it is not expected that one can instantly outperform state-of-the-art large scale TO (millions of elements and constraints), but it strengthens the proof-of-concept to provide qualitative arguments for how the framework allows for future scalability and transferability. 

\paragraph{The intermediate}
The basic benchmark-selection approach should be the minimum to illustrate the potential of a framework in a proof-of-concept manner, in addition to a fair qualitative justification of the method. The next step in proving whether the framework is competitive is to consider whether it allows for changes in the problem definition. This could be included by considering the same benchmark problems as from the basic approach, but now altering or extending the underlying optimisation problem.\\

Admittedly, the basic benchmark problems considered in the TO community have a tendency to be overly simplistic, e.g. by considering minimisation of physical performance subject to a linear volume constraint. This formulation allows for using bi-section algorithms for determining Lagrange multipliers and results in designs that are independent from base material stiffness. Similarly, in learning-based approaches, a formulation with a volume constraint makes training easier and results independent of material properties. For gradient approaches combined with flexible optimisers like the Method of Moving Asymptotes (MMA) (\citealt{Svanberg1987}), switching the objective and constraint functions is straightforward. However, for a learning-based approach, especially aimed at direct design, training becomes increasingly challenging due to varying volume fractions of final designs and the need for quantitative evaluation of the constrained response function.\\

Hence, a challenging test case, with a high degree of industrial and practical relevance, would be to solve volume minimisation subject to compliance constraints. For additional complexity one could also consider solving for different material properties. For extending the problem one can consider additional constraints, such as additional compliance, displacement, local stress or local volume constraints. These changes are all readily implementable within conventional gradient based frameworks and are thus important factors for judging whether the ANN-based method is competitive in a broader more applicable sense.

\paragraph{The advanced}
State-of-the-art TO has come a long way, and is now capable of solving large-scale problems with hundreds of millions of elements, millions of local constraints, a vast variety of alternative physics (a.o. thermofluidics, micro electro mechanical systems) and solved using unstructured meshes on irregular domains. The latter point is of particular interest as the reviewed NN-frameworks for TO exclusively consider regular meshes, which do not guarantee the necessary accuracy for practical applications. The ultimate end-goal for a procedure should be to improve on the performance of conventional methods for such problems or to achieve capabilities for new even more complex or difficult problems. This is, as specified, not necessary for proof-of-concept when presenting new solution frameworks, but this should be the ultimate goal for true scientific progress within the field.

\section{Conclusion}\label{sec:4}

\subsection{Current status for AI in TO}

The recent surge in publications presenting research into exploiting AI-technology for TO is likely motivated by such technologies' positive impact on the field of computer vision. With the desire to eliminate the need for iterative solution procedures in structural optimisation, a large number of neural network models have been suggested, clearly inspired by the success of using deep learning for image segmentation and generation tasks. These direct design models are however found to produce poor designs, be expensive to obtain and be very restricted in terms of the variety of problems and {mesh resolutions} they can handle. Their insufficient performance has, however, not reduced the popularity of the premise of iteration-free TO. It is true that conventional TO consists of a computationally expensive {iteration process}, but the iterative nature in itself is not what makes true large-scale TO impractical. Rather, it is the expense of the inter-iteration computations that pose the real challenge. Therefore, it is postulated that the idea of a direct iteration-free TO should be discarded and that the focus should shift towards alleviating the computational load of the costly components within the iterative process.\\

This literature review includes descriptions of several other application areas, relating to for instance acceleration of the optimisation procedure and post-processing of optimised results. These methods allow for designing models with more specific tasks that have a more realistic possibility of being handled by an approximative mathematical model. Designing viable models for these purposes is proven to be a challenge, as there is still a lack of convincing results presented in the literature. Few of the reviewed articles exhibit promise towards actual scientific progress, but the alternatives for such methods are not exhaustively researched, and greater potential is expected. Therefore, one should definitely not reject all prospects for utilising AI-technology to aid in TO.\\

What is important for overall success within this field, is to adopt a more critical perspective when it comes to evaluating ideas and results both pertaining to work done by oneself and others. Throughout this review it has become clear that there is a lack of understanding both related to model viability and interpretation of results. This is seen both by presenting ML-models with tasks they cannot realistically solve, and insufficient reporting of experimental results. Further, there is a trend of describing the output from research works in terms of what the aim was to achieve, instead of what was achieved. To ensure further progress, it is crucial to present ones research results with full transparency, and to accurately assess the work of others when citing literature.\\

The recommendations made in this review article are meant to instigate the above mentioned needed changes. To summarise the important points of the analysis results, 6 questions to {consider} when working with AI in TO are presented. This list is inspired by the recommendations made by \cite{MarcusDavis2019} for assessing AI research results.
\begin{enumerate}
    \item Disregarding the theoretical expectations, what does the AI system actually achieve?
    \item How general is the approach? (E.g. can it capture all aspects of the problem, or just mimic the provided training data?)
    \item Is there a transparent and thorough presentation of performance? (E.g. is the worst-case performance presented and is the computational gain fairly represented?) 
    \item If it is claimed that an AI system outperforms its conventional counterpart, then on what aspects/measures, and how much better?
    \item How far does successfully solving the presented example instances take us toward achieving AI-based state-of-the-art solution methods?
    \item How robust is the system and what is its generalisation ability? Could it work equally well with other problem characteristics (boundary conditions, loads, etc.), without demanding re-training?
\end{enumerate}
Addressing the above points should at least include listing of breakeven thresholds (\ref{eq:break_thresh}) and generalisation ability (Table \ref{tab:Gen_calc}) or equally transparent alternatives where these measures are not feasible. Further, a fair presentation of solution qualities and comparisons to the most relevant benchmarks, is also expected. Such a proper assessment of method capabilities should be a minimum requirement before publication.\\

Interestingly, it is observed that only {14 out of 111} ML-papers discussed in this review have appeared in the Structural and Multidisciplinary Optimization (SMO) journal. A large majority of papers seems to appear in “non-optimisation” and physics journals, where readers and reviewers may have been less exposed to efficient topology optimisation approaches and, hence, more likely to accept the concept of learning-based methods for inverse design. Even if this review is published in SMO, the ambition and hope is that its message will spread to other journals and scientific societies, such that future research efforts are spent in meaningful ways.

\subsection{Future promise}
As mentioned, the most promising ideas for applications of AI in TO in the current literature, relates to acceleration of the iterative optimisation process or post-processing optimised results for manufacturability. An obvious approach that could offer substantial speed-up is to reduce the number of FEAs needed throughout the optimisation procedure. This could be obtained by either removing a part of the iterative process (\citealt{Kallioras2020}) or replacing the FEA for a polynomial process in a subset of the iterations (\citealt{ChiZhangetal2021, Sasaki2019}). Multi-scale approaches where ML-models are used to map analytical results on a coarse grid to approximate values on a fine grid do also seem viable, but the results should then be appropriately compared to other multi-grid methods.\\

Relating to this, the concept of physics-informed neural networks (PINNs), first introduced by \cite{Raissietal2019}, has lately gained increasing traction and is useful for learning tasks in the presence of physical laws that should be respected. By encoding structured information into the {loss function} of a neural network utilising the principles of PINNs, one could for instance obtain an approximate model describing the solution to a set of partial differential equations that would function as a substitute for FEA. So far experimental results show that this approach requires fewer {data samples} to train a more generalised model. Due to a current lack of combinations with TO this is beyond the scope of this review, but it is believed that this could be an interesting path for further research in the field. This approach may allow for both more efficient and accurate approximations of the governing equations in a topology optimisation problem, as well as modelling of more complex and highly nonlinear mechanical properties. {However, optimisation is known to optimise numerical errors before physics (cf. the checkerboard problem), and therefore caution should be taken when working with potentially inaccurate numerical physics descriptions.}\\

Post-processing of TO-optimised structures utilising ML-methodology is currently an underrepresented application area in the literature that may deserve more attention in the future. Especially, converting the TO-optimised structure representation to a format suited for different manufacturing techniques may have a positive effect on the possibilities for applying TO in real-life product design processes. Also, as shown in \cite{Elingaard2021}, ML has promise for efficient de-homogenisation which strengthens the capabilities of the more efficient homogenisation-based TO method. As such, one can achieve more efficient optimisation procedures also without changing the main optimisation process itself.\\

Further, the structural representation, typically in the form of discretised FE-grids, is a main reason for both the computational time and memory requirements associated with large-scale TO. Therefore, significant improvements could be achieved by appropriate re-parameterisations of the TO-models, where the problem size is reduced in terms of number of design variables or the information needed to sufficiently represent a structure. This is, however, only true provided that FEA is also removed from the usual mesh. As there exists several ML-methods that have proven to be good for feature extraction through down-sampling, this could be another interesting research avenue, related to the PINNs. Note that any such new method should be compared to current {state-of-the-art} model order reduction methods and not to full scale standard TO.\\

Overall, the research into using AI-technology in TO has barely begun. This review has mainly focused on the use of ANNs, but it is believed that many crucial arguments readily translate to other types of ML-models. A select group of applications have so far shown promising results for future development, but most works exhibit unrealistic expectations for what such models can learn. To further advance the use of AI in TO, there is therefore a need for greater knowledge and understanding of existing AI capabilities amongst researchers applying such technology. As new {technology continues} to emerge, some of the recommendations in this paper may change, but the proposed considerations for how to evaluate new frameworks will most probably not.

\begin{appendices}

\section{Fig. \ref{fig:test3} - Upscaling procedure}\label{app:sec1}
\begin{center}
\begin{minipage}{0.9\linewidth}
\begin{lstlisting}[style=Matlab-editor,basicstyle=\scriptsize\ttfamily,keepspaces=true,columns=flexible]
% problem parameters
nelx=60; nely=20; volfrac=0.5; 
penal=3.0; rmin=2.0; ft=1; 
% extract coarse scale design x0 obtained by SIMP
x0=top88(nelx,nely,volfrac,penal,rmin,ft);
nelx=120; nely=40; % fine grid discretisation
% bicubic interpolation for image-resizing to fine grid
x=imresize(x0,[nely,nelx],'method','bicubic');
% apply volume-preserving threshold to obtain discrete structure xt
xt=x; [Y,I]=sort(x(:),'descend');
vt=floor((volfrac-1e-9)*nelx*nely/(1-1e-9));
xt(I(1:vt))=1; xt(I(vt+1:end))=1e-9;
\end{lstlisting}
\end{minipage}
\end{center}




\end{appendices}

\section*{Declarations}

\textbf{Conflict of interest:} The authors state that there is no conflict of interest.

\noindent \textbf{Funding:} The authors acknowledge funding from the Villum Fonden through the Villum Investigator Project ``InnoTop''. 
%The information provided in this paper is the sole opinion of the authors and does not necessarily reflect the view of the sponsoring agencies.

\noindent\textbf{Replication of results:} The results presented in this review can be reproduced by the provided descriptions of the experiments and the corresponding referenced open-source codes. 
% s All results in this paper are generated by codes and data from source references. Readers are encouraged to download papers and codes of interest from original publications

%%%%%%%%%%%% Supplementary Methods %%%%%%%%%%%%
%\footnotesize
%\section*{Methods}

%%%%%%%%%%%%% Acknowledgements %%%%%%%%%%%%%
%\footnotesize
%\section*{Acknowledgements}

%%%%%%%%%%%%%%   Bibliography   %%%%%%%%%%%%%%
\normalsize
\begingroup
\raggedright

\bibliography{references}
\endgroup
%%%%%%%%%%%%  Supplementary Figures  %%%%%%%%%%%%
%\clearpage

%%%%%%%%%%%%%%%%   End   %%%%%%%%%%%%%%%%
%\end{multicols}  % Method B for two-column formatting (doesn't play well with line numbers), comment out if using method A
\end{document}